\PassOptionsToPackage{unicode}{hyperref}
\PassOptionsToPackage{hyphens}{url}
\documentclass[
  12pt,
]{article}
\usepackage{xcolor}
\usepackage[margin=1in]{geometry}
\usepackage{amsmath,amssymb}
\usepackage{iftex}
\ifPDFTeX
  \usepackage[T1]{fontenc}
  \usepackage[utf8]{inputenc}
  \usepackage{textcomp} 
\else 
  \usepackage{unicode-math} 
  \defaultfontfeatures{Scale=MatchLowercase}
  \defaultfontfeatures[\rmfamily]{Ligatures=TeX,Scale=1}
\fi
\usepackage{lmodern}
\ifPDFTeX\else
\fi
\IfFileExists{upquote.sty}{\usepackage{upquote}}{}
\IfFileExists{microtype.sty}{
  \usepackage[]{microtype}
  \UseMicrotypeSet[protrusion]{basicmath} 
}{}
\makeatletter
\@ifundefined{KOMAClassName}{
  \IfFileExists{parskip.sty}{%
    \usepackage{parskip}
  }{
    \setlength{\parindent}{0pt}
    \setlength{\parskip}{6pt plus 2pt minus 1pt}}
}{
  \KOMAoptions{parskip=half}}
\makeatother
\usepackage{color}
\usepackage{fancyvrb}

\DefineVerbatimEnvironment{Highlighting}{Verbatim}{commandchars=\\\{\}}
\usepackage{framed}
\definecolor{shadecolor}{RGB}{248,248,248}
\newenvironment{Shaded}{\begin{snugshade}}{\end{snugshade}}

\newcommand{\BuiltInTok}[1]{#1}

\newcommand{\ControlFlowTok}[1]{\textcolor[rgb]{0.13,0.29,0.53}{\textbf{#1}}}

\newcommand{\DecValTok}[1]{\textcolor[rgb]{0.00,0.00,0.81}{#1}}

\newcommand{\FloatTok}[1]{\textcolor[rgb]{0.00,0.00,0.81}{#1}}

\newcommand{\ImportTok}[1]{#1}

\newcommand{\KeywordTok}[1]{\textcolor[rgb]{0.13,0.29,0.53}{\textbf{#1}}}
\newcommand{\NormalTok}[1]{#1}
\newcommand{\OperatorTok}[1]{\textcolor[rgb]{0.81,0.36,0.00}{\textbf{#1}}}

\newcommand{\PreprocessorTok}[1]{\textcolor[rgb]{0.56,0.35,0.01}{\textit{#1}}}

\newcommand{\SpecialCharTok}[1]{\textcolor[rgb]{0.81,0.36,0.00}{\textbf{#1}}}
\newcommand{\SpecialStringTok}[1]{\textcolor[rgb]{0.31,0.60,0.02}{#1}}
\newcommand{\StringTok}[1]{\textcolor[rgb]{0.31,0.60,0.02}{#1}}
\newcommand{\VariableTok}[1]{\textcolor[rgb]{0.00,0.00,0.00}{#1}}
\newcommand{\VerbatimStringTok}[1]{\textcolor[rgb]{0.31,0.60,0.02}{#1}}

\usepackage{longtable,booktabs,array}
\usepackage{calc} 
\usepackage{etoolbox}
\makeatletter
\patchcmd\longtable{\par}{\if@noskipsec\mbox{}\fi\par}{}{}
\makeatother
\IfFileExists{footnotehyper.sty}{\usepackage{footnotehyper}}{\usepackage{footnote}}
\makesavenoteenv{longtable}
\providecommand{\tightlist}{%
  \setlength{\itemsep}{0pt}\setlength{\parskip}{0pt}}
\usepackage{graphicx}
\usepackage{subcaption}
\usepackage{booktabs}
\usepackage{longtable}
\usepackage{float}
\usepackage{hyperref}
\usepackage{xcolor}
\usepackage{microtype}
\usepackage{geometry}
\fvset{fontsize=\footnotesize}
\AtBeginEnvironment{longtable}{\footnotesize}

\hypersetup{
  colorlinks=true,
  linkcolor=blue,
  citecolor=blue,
  urlcolor=blue
}

\usepackage{bookmark}
\IfFileExists{xurl.sty}{\usepackage{xurl}}{} 
\hypersetup{
  pdftitle={Pandas for Reproducible Data Analysis: From Spreadsheets to Research-Grade Python Workflows},
  hidelinks,
  pdfcreator={LaTeX via pandoc}}

\title{Pandas for Reproducible Data Analysis: From Spreadsheets to
Research-Grade Python Workflows}
\author{
Sidney Shapiro\\
Dhillon School of Business, University of Lethbridge\\
Lethbridge, Alberta, Canada
\and
Daniel Pearson\\
Dhillon School of Business, University of Lethbridge\\
Lethbridge, Alberta, Canada
\and
Emiliano Sebastian Gonzalez Venegas\\
Universidad de Guadalajara\\
Guadalajara, Jalisco, Mexico
}
\date{}

\begin{document}
\maketitle

\section{Abstract}\label{abstract}

Spreadsheet-heavy analytical work remains the default in business
analytics, operations reporting, and applied research, yet workbooks
that grow through formulas, manual edits, and copy-paste refresh are
difficult to audit, reproduce, and govern at scale. When tabular work
requires repeatability, validation, version control, automated refresh,
or integration with statistics and machine learning, analysts need a
transformation layer that preserves familiar table concepts while making
assumptions explicit. The paper treats the Python \texttt{pandas}
library as that layer: a practical bridge between spreadsheet practice
and research-grade workflows, not a wholesale replacement for Excel. The
paper contributes an Excel-to-pandas migration mapping, a taxonomy of
nine workflow categories, seven end-to-end examples drawn from business
analytics and applied research, a failure-mode catalog, and reusable
code recipes for governed tabular work. \texttt{pandas} is most useful
when tabular analysis must be repeatable, auditable, and defensible,
while Excel can remain a familiar input and output interface for
stakeholders who need workbooks.

\textbf{Keywords:} pandas, Python, data analysis, DataFrame, spreadsheet
migration, Excel, business analytics, reproducible research, data
validation, data cleaning, reporting, machine learning preparation

\section{1. Introduction}\label{introduction}

Data analysis work is increasingly split across two worlds. The first is
the spreadsheet world: immediately accessible, visual, flexible, and
deeply embedded in business practice. The second is the programmatic
analysis world: scriptable, repeatable, testable, and better suited to
data pipelines, collaborative research, and production systems. Many
researchers and practitioners do not begin in the second world. They
arrive there after building operational workbooks, reconciling monthly
reports, cleaning survey exports, or maintaining decision models whose
complexity eventually exceeds the practical limits of manual spreadsheet
work. The problem is not spreadsheet use itself. The problem is that
consequential tabular work---monthly revenue close, workforce planning,
compliance tracking, survey research, and pricing models---often remains
trapped in opaque workbooks when it should be governed like other
critical systems. Spreadsheet formulas answer questions in place; they
rarely record how an answer was produced, what assumptions were checked,
or which version of the data was used. When the same report must be
refreshed monthly, joined to reference tables, validated against control
totals, or handed to a statistician or modeler, manual workbook logic
becomes a reproducibility and governance bottleneck.

Business analytics curricula and workplace training often teach Excel
deeply but introduce programmatic data work only later in the
curriculum. Applied researchers may learn Python or R for modeling while
still receiving collaborator data as messy \texttt{.xlsx} files. Data
engineering literature, meanwhile, emphasizes warehouses, orchestration,
and schema-first pipelines that can feel distant from the day-to-day
work of reconciling a regional sales workbook. The Python
\texttt{pandas} library, introduced as a DataFrame-centered tool for
statistical computing in Python {[}2{]}, sits in that gap. Its
\texttt{DataFrame} is close enough to a worksheet to lower the adoption
barrier, yet expressive enough to encode validation, joins, reshaping,
and reporting as reusable code. A workbook formula answers a question in
place; a \texttt{pandas} pipeline records how the answer was produced.
That distinction matters for scientific reproducibility {[}10, 11{]},
auditability, regulatory reporting, and ordinary business continuity.
Modern \texttt{pandas} also integrates with NumPy, visualization
libraries, SQL engines, and scikit-learn {[}3, 4, 5{]}, so the same
prepared table can support exploration, publication graphics, and
modeling without re-entering data.

This paper documents that transition for spreadsheet-experienced
analysts, instructors, and applied researchers. Its contribution is
practical rather than algorithmic: it provides an Excel-to-pandas
migration mapping, a workflow taxonomy from ingestion through
reproducibility, a catalog of common failure modes, and reusable code
recipes for governed tabular work. The focus is applied. We treat
\texttt{pandas} as a working tool in research labs, finance teams,
operations groups, public sector reporting, survey research, and data
science, using examples that include workbook import, VLOOKUP-style
joins, pivot-style summaries, currency and percent cleaning, monthly
dashboards, and Excel export for stakeholders. Figure 1 situates
\texttt{pandas} in the analytical stack; Tables 1--3 and Figures 2--3
provide the conceptual scaffolding used throughout the paper.

\section{2. Why Pandas Matters After
Spreadsheets}\label{why-pandas-matters-after-spreadsheets}

Spreadsheets remain valuable. They are excellent for small exploratory
tasks, simple models, data entry, ad hoc review, and communication with
nontechnical stakeholders. The problem is not spreadsheet use itself.
The problem is relying on opaque workbooks as the permanent system of
record for work that needs repeatability, peer review, provenance,
scale, or automated refresh.

Spreadsheet-based workflows often acquire risk gradually:

\begin{itemize}
\tightlist
\item
  formulas are copied across ranges and later edited inconsistently;
\item
  hidden rows, filters, manual sorts, and merged cells affect
  interpretation;
\item
  external links break or silently reference stale files;
\item
  data types are inferred visually rather than declared;
\item
  business rules live in cell formulas rather than version-controlled
  code;
\item
  final reports are produced through manual copy-and-paste operations;
\item
  reviewers can inspect final numbers but not always the full
  transformation path.
\end{itemize}

Spreadsheet error research has repeatedly noted that spreadsheet errors
are common and difficult to detect in consequential workbooks {[}7, 8,
16{]}. In business settings, this is not merely a technical problem. It
is a governance problem. A monthly revenue workbook, workforce planning
file, compliance tracker, or pricing model may become critical
infrastructure without the controls normally expected of critical
systems.

\texttt{pandas} offers a pragmatic response. It does not remove Excel
from the workflow. Instead, it can move data cleaning, transformation,
reconciliation, and metric generation into code, while still reading
from and writing to Excel when Excel is the interface stakeholders
expect. Framed this way, \texttt{pandas} is a \textbf{reproducible
transformation layer} between raw data and analytical decisions (Figure
1).

\textbf{Figure 1.} Pandas as a reproducible transformation layer between
raw data sources and analytical outputs. Excel and databases remain
valid inputs and outputs; pandas encodes the governed middle stage.

\begin{center}
\includegraphics[width=0.96\linewidth,height=0.62\textheight,keepaspectratio]{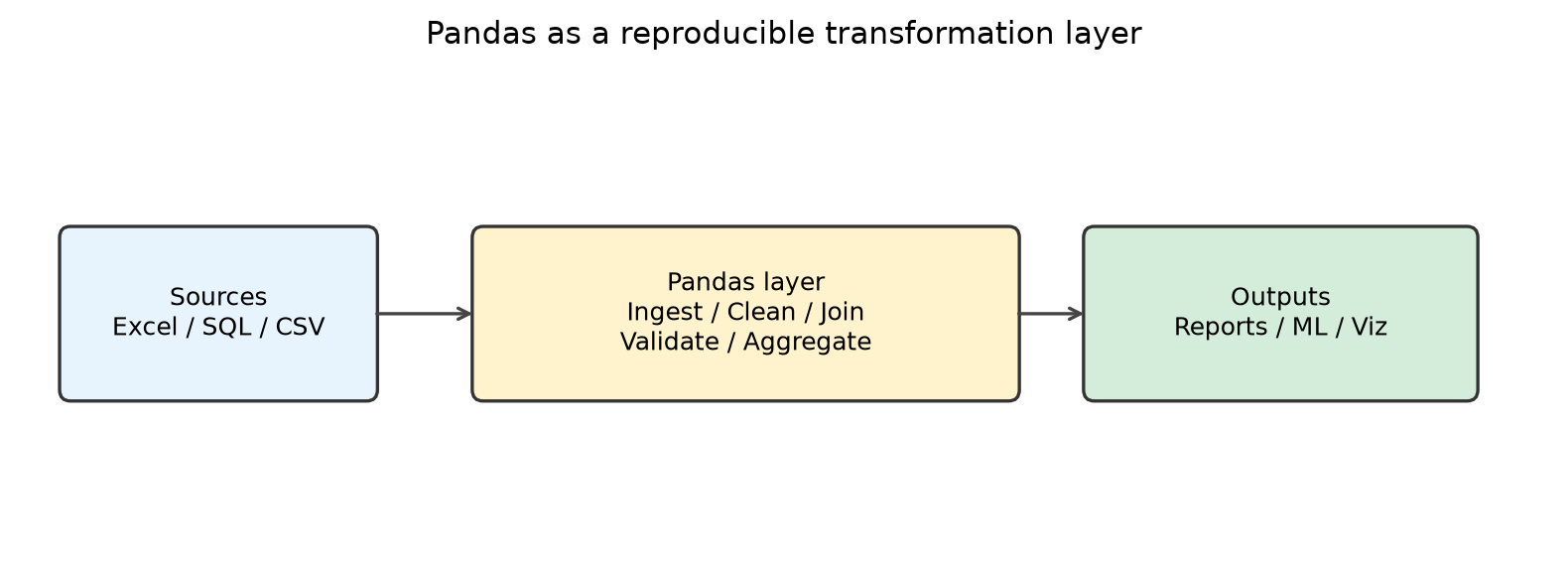}
\end{center}

Table 1 links common spreadsheet governance risks to the controls that
\texttt{pandas} workflows can encode explicitly.

\textbf{Table 1.} Spreadsheet governance risks and pandas mitigations

{\def\LTcaptype{none} 
\begin{longtable}[]{@{}
  >{\raggedright\arraybackslash}p{(\linewidth - 4\tabcolsep) * \real{0.3333}}
  >{\raggedright\arraybackslash}p{(\linewidth - 4\tabcolsep) * \real{0.3333}}
  >{\raggedright\arraybackslash}p{(\linewidth - 4\tabcolsep) * \real{0.3333}}@{}}
\toprule\noalign{}
\begin{minipage}[b]{\linewidth}\raggedright
Risk in spreadsheet-heavy work
\end{minipage} & \begin{minipage}[b]{\linewidth}\raggedright
Why it matters
\end{minipage} & \begin{minipage}[b]{\linewidth}\raggedright
Pandas mitigation
\end{minipage} \\
\midrule\noalign{}
\endhead
\bottomrule\noalign{}
\endlastfoot
Formula drift across copied ranges & Inconsistent business rules & Named
column operations in version-controlled scripts \\
Hidden rows, filters, subtotals & Misread grain and double-counting &
Explicit row filters and subtotal removal in code \\
Untyped identifiers & Broken joins, lost leading zeros & \texttt{dtype=}
declarations at read time \\
Opaque VLOOKUP failures & Silent nulls and inflated totals &
\texttt{merge()} with \texttt{indicator=} and \texttt{validate=} \\
Manual report assembly & Stale or unreconciled numbers & Assertions
before \texttt{to\_excel()} \\
No transformation record & Weak audit and reproducibility & Scripts,
logs, and immutable \texttt{data/raw/} \\
\end{longtable}
}

\subsection{2.1 Mapping Excel Concepts to
Pandas}\label{mapping-excel-concepts-to-pandas}

Figure 2 contrasts the mental model shift from cell formulas to column
rules. Table 2 summarizes the operational mapping used in the examples
that follow.

\textbf{Figure 2.} Cell-oriented spreadsheet logic versus
column-oriented pandas transformations. The same business rule applies
to every row, but the pandas version is inspectable as code.

\begin{center}
\includegraphics[width=0.96\linewidth,height=0.62\textheight,keepaspectratio]{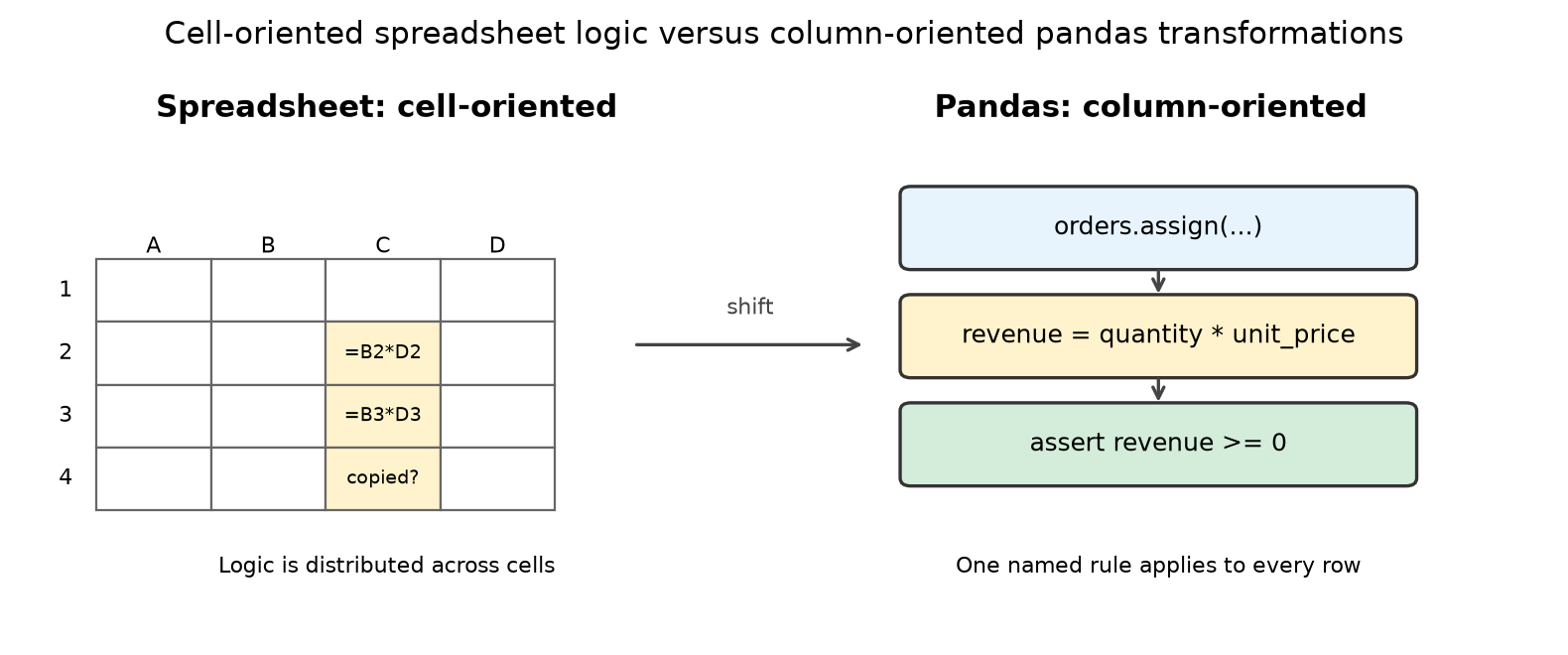}
\end{center}

\textbf{Table 2.} Mapping Excel concepts to pandas equivalents

{\def\LTcaptype{none} 
\begin{longtable}[]{@{}
  >{\raggedright\arraybackslash}p{(\linewidth - 4\tabcolsep) * \real{0.3333}}
  >{\raggedright\arraybackslash}p{(\linewidth - 4\tabcolsep) * \real{0.3333}}
  >{\raggedright\arraybackslash}p{(\linewidth - 4\tabcolsep) * \real{0.3333}}@{}}
\toprule\noalign{}
\begin{minipage}[b]{\linewidth}\raggedright
Spreadsheet concept
\end{minipage} & \begin{minipage}[b]{\linewidth}\raggedright
Pandas equivalent
\end{minipage} & \begin{minipage}[b]{\linewidth}\raggedright
Typical reason to switch
\end{minipage} \\
\midrule\noalign{}
\endhead
\bottomrule\noalign{}
\endlastfoot
Worksheet & \texttt{DataFrame} & Work with named columns and typed
values. \\
Column formula & Vectorized expression or \texttt{.assign()} & Apply one
rule consistently to all rows. \\
VLOOKUP/XLOOKUP & \texttt{merge()} with \texttt{validate=} & Join tables
and detect relationship errors. \\
Pivot table & \texttt{groupby()}, \texttt{pivot\_table()} & Produce
repeatable aggregations. \\
Filter & Boolean mask, \texttt{.query()} & Make row selection explicit
and reusable. \\
Text-to-columns & \texttt{.str} methods, \texttt{extract()},
\texttt{split()} & Clean text with documented rules. \\
Remove duplicates & \texttt{drop\_duplicates()} & Define duplicate keys
explicitly. \\
Conditional formatting checks & assertions and validation tables & Fail
fast when business rules are violated. \\
Manual report workbook & \texttt{ExcelWriter}, HTML, dashboards &
Generate outputs from the same source code. \\
Copy-paste refresh & parameterized script or scheduled job & Reduce
manual intervention and stale data. \\
\end{longtable}
}

The conceptual move is from cell-level editing to table-level
transformation. This is the most important adjustment for Excel users. A
spreadsheet user may ask, ``What formula should I put in this cell?'' A
\texttt{pandas} user usually asks, ``What rule should create this column
for every applicable row, and how do I verify the result?''

\section{3. A Taxonomy of Pandas Workflow
Patterns}\label{a-taxonomy-of-pandas-workflow-patterns}

Applied \texttt{pandas} work rarely consists of isolated function calls.
In practice, analysts chain reusable workflow patterns. Table 3
summarizes nine categories that appear repeatedly in business analytics
and applied research. Each pattern answers a different governance
question: where data entered, how it was cleaned, whether joins are
trustworthy, and how outputs can be reproduced. Figure 3 shows how these
categories typically chain in a monthly reporting pipeline.

\textbf{Table 3.} Taxonomy of pandas workflow patterns

{\def\LTcaptype{none} 
\begin{longtable}[]{@{}
  >{\raggedright\arraybackslash}p{(\linewidth - 8\tabcolsep) * \real{0.2000}}
  >{\raggedright\arraybackslash}p{(\linewidth - 8\tabcolsep) * \real{0.2000}}
  >{\raggedright\arraybackslash}p{(\linewidth - 8\tabcolsep) * \real{0.2000}}
  >{\raggedright\arraybackslash}p{(\linewidth - 8\tabcolsep) * \real{0.2000}}
  >{\raggedright\arraybackslash}p{(\linewidth - 8\tabcolsep) * \real{0.2000}}@{}}
\toprule\noalign{}
\begin{minipage}[b]{\linewidth}\raggedright
\#
\end{minipage} & \begin{minipage}[b]{\linewidth}\raggedright
Category
\end{minipage} & \begin{minipage}[b]{\linewidth}\raggedright
Primary question
\end{minipage} & \begin{minipage}[b]{\linewidth}\raggedright
Key pandas tools
\end{minipage} & \begin{minipage}[b]{\linewidth}\raggedright
Typical outputs
\end{minipage} \\
\midrule\noalign{}
\endhead
\bottomrule\noalign{}
\endlastfoot
1 & Ingestion & Did we read the source faithfully? &
\texttt{read\_excel}, \texttt{read\_csv}, \texttt{read\_sql} & Typed
\texttt{DataFrame}, ingest log \\
2 & Cleaning & Are values standardized? & \texttt{rename},
\texttt{assign}, \texttt{dropna} & Clean table, dropped-row report \\
3 & Enrichment & Do keys match reference data? & \texttt{merge},
\texttt{validate=} & Joined table, unmatched-key report \\
4 & Reshaping & Is the grain correct? & \texttt{melt},
\texttt{pivot\_table} & Long/tidy or wide report table \\
5 & Aggregation & What metrics are needed? & \texttt{groupby},
\texttt{agg} & Summary metrics, pivot tables \\
6 & Validation & Do rules and totals hold? & \texttt{assert},
reconciliation & QA tables, failure logs \\
7 & Reporting & How do stakeholders consume results? &
\texttt{ExcelWriter}, \texttt{to\_html} & Workbooks, dashboards \\
8 & Modeling preparation & Are features leakage-free? & sklearn
pipelines & Model-ready data \\
9 & Reproducibility & Can this rerun next month? & scripts, env files &
Project layout \\
\end{longtable}
}

\textbf{Figure 3.} Sequential flow of pandas workflow pattern categories
in a typical governed analytics pipeline.

\begin{center}
\includegraphics[width=0.96\linewidth,height=0.62\textheight,keepaspectratio]{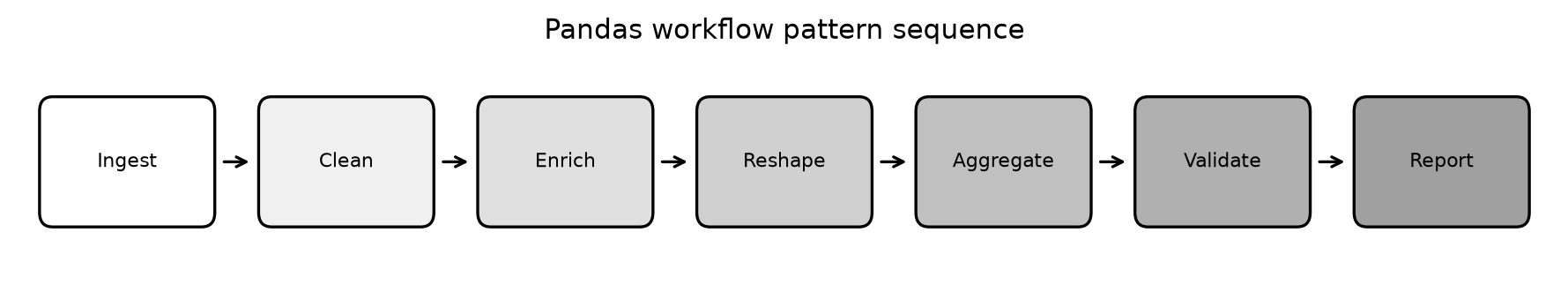}
\end{center}

\subsection{3.1 Ingestion Workflows}\label{ingestion-workflows}

Ingestion workflows read Excel, CSV, databases, or APIs while preserving
identifiers and declaring dtypes early.

\begin{Shaded}
\begin{Highlighting}[]
\ImportTok{import}\NormalTok{ pandas }\ImportTok{as}\NormalTok{ pd}
\ImportTok{from}\NormalTok{ pathlib }\ImportTok{import}\NormalTok{ Path}

\NormalTok{sales }\OperatorTok{=}\NormalTok{ pd.read\_excel(}
\NormalTok{    Path(}\StringTok{"data/raw/monthly\_sales.xlsx"}\NormalTok{),}
\NormalTok{    sheet\_name}\OperatorTok{=}\StringTok{"Sales"}\NormalTok{,}
\NormalTok{    dtype}\OperatorTok{=}\NormalTok{\{}\StringTok{"order\_id"}\NormalTok{: }\StringTok{"string"}\NormalTok{, }\StringTok{"customer\_id"}\NormalTok{: }\StringTok{"string"}\NormalTok{\},}
\NormalTok{    parse\_dates}\OperatorTok{=}\NormalTok{[}\StringTok{"order\_date"}\NormalTok{],}
\NormalTok{)}
\BuiltInTok{print}\NormalTok{(}\SpecialStringTok{f"ingested }\SpecialCharTok{\{}\BuiltInTok{len}\NormalTok{(sales)}\SpecialCharTok{:,\}}\SpecialStringTok{ sales rows"}\NormalTok{)}
\end{Highlighting}
\end{Shaded}

\subsection{3.2 Cleaning Workflows}\label{cleaning-workflows}

Cleaning workflows standardize names, parse dates, strip currency
symbols, and remove subtotal rows.

\begin{Shaded}
\begin{Highlighting}[]
\NormalTok{clean }\OperatorTok{=}\NormalTok{ (}
\NormalTok{  sales.rename(columns}\OperatorTok{=}\KeywordTok{lambda}\NormalTok{ c: c.strip().lower().replace(}\StringTok{" "}\NormalTok{, }\StringTok{"\_"}\NormalTok{))}
\NormalTok{  .loc[}\KeywordTok{lambda}\NormalTok{ d: d[}\StringTok{"order\_id"}\NormalTok{].notna() }\OperatorTok{\&} \OperatorTok{\textasciitilde{}}\NormalTok{d[}\StringTok{"order\_id"}\NormalTok{].}\BuiltInTok{str}\NormalTok{.lower().eq(}\StringTok{"total"}\NormalTok{)]}
\NormalTok{  .assign(order\_date}\OperatorTok{=}\KeywordTok{lambda}\NormalTok{ d: pd.to\_datetime(d[}\StringTok{"order\_date"}\NormalTok{], errors}\OperatorTok{=}\StringTok{"coerce"}\NormalTok{))}
\NormalTok{)}
\end{Highlighting}
\end{Shaded}

\subsection{3.3 Enrichment Workflows}\label{enrichment-workflows}

Enrichment workflows join lookup tables with explicit cardinality
checks.

\begin{Shaded}
\begin{Highlighting}[]
\NormalTok{enriched }\OperatorTok{=}\NormalTok{ orders.merge(}
\NormalTok{    customers,}
\NormalTok{    on}\OperatorTok{=}\StringTok{"customer\_id"}\NormalTok{,}
\NormalTok{    how}\OperatorTok{=}\StringTok{"left"}\NormalTok{,}
\NormalTok{    validate}\OperatorTok{=}\StringTok{"m:1"}\NormalTok{,}
\NormalTok{    indicator}\OperatorTok{=}\VariableTok{True}\NormalTok{,}
\NormalTok{)}
\NormalTok{unmatched }\OperatorTok{=}\NormalTok{ enriched.loc[enriched[}\StringTok{"\_merge"}\NormalTok{].eq(}\StringTok{"left\_only"}\NormalTok{), }\StringTok{"customer\_id"}\NormalTok{].unique()}
\end{Highlighting}
\end{Shaded}

\subsection{3.4 Reshaping Workflows}\label{reshaping-workflows}

Reshaping workflows convert wide exports to tidy long format or build
wide management reports.

\begin{Shaded}
\begin{Highlighting}[]
\NormalTok{long\_survey }\OperatorTok{=}\NormalTok{ wide\_survey.melt(}
\NormalTok{    id\_vars}\OperatorTok{=}\StringTok{"respondent\_id"}\NormalTok{,}
\NormalTok{    var\_name}\OperatorTok{=}\StringTok{"question"}\NormalTok{,}
\NormalTok{    value\_name}\OperatorTok{=}\StringTok{"response"}\NormalTok{,}
\NormalTok{)}
\end{Highlighting}
\end{Shaded}

\subsection{3.5 Aggregation Workflows}\label{aggregation-workflows}

Aggregation workflows compute monthly metrics, customer summaries, or
survey department means.

\begin{Shaded}
\begin{Highlighting}[]
\NormalTok{monthly }\OperatorTok{=}\NormalTok{ enriched.groupby([}\StringTok{"month"}\NormalTok{, }\StringTok{"region"}\NormalTok{], observed}\OperatorTok{=}\VariableTok{True}\NormalTok{).agg(revenue}\OperatorTok{=}\NormalTok{(}\StringTok{"revenue"}\NormalTok{, }\StringTok{"sum"}\NormalTok{))}
\end{Highlighting}
\end{Shaded}

\subsection{3.6 Validation Workflows}\label{validation-workflows}

Validation workflows encode business rules as assertions and
reconciliation tables.

\begin{Shaded}
\begin{Highlighting}[]
\ControlFlowTok{assert}\NormalTok{ np.isclose(detail[}\StringTok{"revenue"}\NormalTok{].}\BuiltInTok{sum}\NormalTok{(), summary[}\StringTok{"revenue"}\NormalTok{].}\BuiltInTok{sum}\NormalTok{())}
\end{Highlighting}
\end{Shaded}

\subsection{3.7 Reporting Workflows}\label{reporting-workflows}

Reporting workflows export validated tables to Excel or HTML after
computation is complete.

\begin{Shaded}
\begin{Highlighting}[]
\ControlFlowTok{with}\NormalTok{ pd.ExcelWriter(}\StringTok{"reports/monthly\_metrics.xlsx"}\NormalTok{, engine}\OperatorTok{=}\StringTok{"xlsxwriter"}\NormalTok{) }\ImportTok{as}\NormalTok{ writer:}
\NormalTok{    monthly.to\_excel(writer, sheet\_name}\OperatorTok{=}\StringTok{"Metrics"}\NormalTok{, index}\OperatorTok{=}\VariableTok{False}\NormalTok{)}
\end{Highlighting}
\end{Shaded}

\subsection{3.8 Modeling Preparation
Workflows}\label{modeling-preparation-workflows}

Modeling preparation keeps column meaning explicit and moves
preprocessing into sklearn pipelines.

\begin{Shaded}
\begin{Highlighting}[]
\NormalTok{preprocess }\OperatorTok{=}\NormalTok{ ColumnTransformer([(}\StringTok{"num"}\NormalTok{, numeric\_pipe, num\_cols), (}\StringTok{"cat"}\NormalTok{, cat\_pipe, cat\_cols)])}
\end{Highlighting}
\end{Shaded}

\subsection{3.9 Reproducibility
Workflows}\label{reproducibility-workflows}

Reproducibility workflows separate raw from processed data, pin
dependencies, and prefer scripts for repeated jobs.

\begin{Shaded}
\begin{Highlighting}[]
\NormalTok{clean.to\_parquet(}\StringTok{"data/processed/orders.parquet"}\NormalTok{, index}\OperatorTok{=}\VariableTok{False}\NormalTok{)}
\end{Highlighting}
\end{Shaded}

\section{4. Core Programming Model}\label{core-programming-model}

The central objects in \texttt{pandas} are \texttt{Series} and
\texttt{DataFrame}. A \texttt{Series} is a one-dimensional labeled
array. A \texttt{DataFrame} is a two-dimensional table whose columns may
have different data types. In business terms, a \texttt{DataFrame}
resembles a worksheet or database table, but with operations designed
for code.

\begin{Shaded}
\begin{Highlighting}[]
\ImportTok{import}\NormalTok{ pandas }\ImportTok{as}\NormalTok{ pd}

\NormalTok{orders }\OperatorTok{=}\NormalTok{ pd.DataFrame(\{}
    \StringTok{"order\_id"}\NormalTok{: [}\DecValTok{1001}\NormalTok{, }\DecValTok{1002}\NormalTok{, }\DecValTok{1003}\NormalTok{, }\DecValTok{1004}\NormalTok{],}
    \StringTok{"customer\_id"}\NormalTok{: [}\StringTok{"C001"}\NormalTok{, }\StringTok{"C002"}\NormalTok{, }\StringTok{"C001"}\NormalTok{, }\StringTok{"C003"}\NormalTok{],}
    \StringTok{"order\_date"}\NormalTok{: pd.to\_datetime([}\StringTok{"2025{-}01{-}05"}\NormalTok{, }\StringTok{"2025{-}01{-}07"}\NormalTok{, }\StringTok{"2025{-}02{-}03"}\NormalTok{, }\StringTok{"2025{-}02{-}20"}\NormalTok{]),}
    \StringTok{"region"}\NormalTok{: [}\StringTok{"West"}\NormalTok{, }\StringTok{"East"}\NormalTok{, }\StringTok{"West"}\NormalTok{, }\StringTok{"Central"}\NormalTok{],}
    \StringTok{"quantity"}\NormalTok{: [}\DecValTok{3}\NormalTok{, }\DecValTok{5}\NormalTok{, }\DecValTok{2}\NormalTok{, }\DecValTok{8}\NormalTok{],}
    \StringTok{"unit\_price"}\NormalTok{: [}\FloatTok{120.00}\NormalTok{, }\FloatTok{45.50}\NormalTok{, }\FloatTok{120.00}\NormalTok{, }\FloatTok{15.25}\NormalTok{],}
\NormalTok{\})}

\NormalTok{orders }\OperatorTok{=}\NormalTok{ orders.assign(}
\NormalTok{    revenue}\OperatorTok{=}\KeywordTok{lambda}\NormalTok{ d: d[}\StringTok{"quantity"}\NormalTok{] }\OperatorTok{*}\NormalTok{ d[}\StringTok{"unit\_price"}\NormalTok{],}
\NormalTok{    month}\OperatorTok{=}\KeywordTok{lambda}\NormalTok{ d: d[}\StringTok{"order\_date"}\NormalTok{].dt.to\_period(}\StringTok{"M"}\NormalTok{),}
\NormalTok{)}

\NormalTok{monthly\_revenue }\OperatorTok{=}\NormalTok{ (}
\NormalTok{    orders}
\NormalTok{    .groupby([}\StringTok{"month"}\NormalTok{, }\StringTok{"region"}\NormalTok{], observed}\OperatorTok{=}\VariableTok{True}\NormalTok{)}
\NormalTok{    .agg(}
\NormalTok{        revenue}\OperatorTok{=}\NormalTok{(}\StringTok{"revenue"}\NormalTok{, }\StringTok{"sum"}\NormalTok{),}
\NormalTok{        orders}\OperatorTok{=}\NormalTok{(}\StringTok{"order\_id"}\NormalTok{, }\StringTok{"nunique"}\NormalTok{),}
\NormalTok{        units}\OperatorTok{=}\NormalTok{(}\StringTok{"quantity"}\NormalTok{, }\StringTok{"sum"}\NormalTok{),}
\NormalTok{    )}
\NormalTok{    .reset\_index()}
\NormalTok{)}
\end{Highlighting}
\end{Shaded}

This short example demonstrates several design features. Columns have
names. Dates are actual datetimes rather than formatted text. New
variables are derived by column operations rather than copied formulas.
Aggregation is expressed as a reusable transformation. The code can be
reviewed, tested, committed to version control, and rerun on next
month's data.

\subsection{4.1 Data Types and Missing
Values}\label{data-types-and-missing-values}

Data type management is one of the most important differences between
spreadsheets and \texttt{pandas}. In Excel, values are often interpreted
through formatting. In \texttt{pandas}, a date, string, integer,
decimal, category, and missing value have distinct computational
behavior.

\begin{Shaded}
\begin{Highlighting}[]
\NormalTok{raw }\OperatorTok{=}\NormalTok{ pd.DataFrame(\{}
    \StringTok{" Order ID "}\NormalTok{: [}\StringTok{"1001"}\NormalTok{, }\StringTok{"1002"}\NormalTok{, }\StringTok{""}\NormalTok{, }\StringTok{"Total"}\NormalTok{],}
    \StringTok{"Order Date"}\NormalTok{: [}\StringTok{"1/5/2025"}\NormalTok{, }\StringTok{"2025{-}01{-}06"}\NormalTok{, }\StringTok{"2025{-}01{-}07"}\NormalTok{, }\StringTok{""}\NormalTok{],}
    \StringTok{"Customer"}\NormalTok{: [}\StringTok{"Acme Inc."}\NormalTok{, }\StringTok{"Northwind"}\NormalTok{, }\StringTok{"Northwind"}\NormalTok{, }\StringTok{""}\NormalTok{],}
    \StringTok{"Revenue"}\NormalTok{: [}\StringTok{"$1,250.50"}\NormalTok{, }\StringTok{"$840.00"}\NormalTok{, }\VariableTok{None}\NormalTok{, }\StringTok{"$2,090.50"}\NormalTok{],}
    \StringTok{"Margin \%"}\NormalTok{: [}\StringTok{"32\%"}\NormalTok{, }\StringTok{"28.5\%"}\NormalTok{, }\StringTok{""}\NormalTok{, }\StringTok{""}\NormalTok{],}
    \StringTok{"Region"}\NormalTok{: [}\StringTok{"West"}\NormalTok{, }\StringTok{"East"}\NormalTok{, }\StringTok{"East"}\NormalTok{, }\StringTok{""}\NormalTok{],}
\NormalTok{\})}

\NormalTok{clean }\OperatorTok{=}\NormalTok{ (}
\NormalTok{    raw}
\NormalTok{    .rename(columns}\OperatorTok{=}\KeywordTok{lambda}\NormalTok{ c: c.strip().lower().replace(}\StringTok{" "}\NormalTok{, }\StringTok{"\_"}\NormalTok{).replace(}\StringTok{"\%"}\NormalTok{, }\StringTok{"pct"}\NormalTok{))}
\NormalTok{    .replace(}\VerbatimStringTok{r"}\DecValTok{\^{}\textbackslash{}s}\OperatorTok{*}\DecValTok{$}\VerbatimStringTok{"}\NormalTok{, pd.NA, regex}\OperatorTok{=}\VariableTok{True}\NormalTok{)}
\NormalTok{    .dropna(subset}\OperatorTok{=}\NormalTok{[}\StringTok{"order\_id"}\NormalTok{])}
\NormalTok{    .loc[}\KeywordTok{lambda}\NormalTok{ d: d[}\StringTok{"order\_id"}\NormalTok{].}\BuiltInTok{str}\NormalTok{.lower().ne(}\StringTok{"total"}\NormalTok{)]}
\NormalTok{    .assign(}
\NormalTok{        order\_id}\OperatorTok{=}\KeywordTok{lambda}\NormalTok{ d: d[}\StringTok{"order\_id"}\NormalTok{].astype(}\StringTok{"string"}\NormalTok{),}
\NormalTok{        order\_date}\OperatorTok{=}\KeywordTok{lambda}\NormalTok{ d: pd.to\_datetime(d[}\StringTok{"order\_date"}\NormalTok{], errors}\OperatorTok{=}\StringTok{"coerce"}\NormalTok{),}
\NormalTok{        revenue}\OperatorTok{=}\KeywordTok{lambda}\NormalTok{ d: (}
\NormalTok{            d[}\StringTok{"revenue"}\NormalTok{]}
\NormalTok{            .astype(}\StringTok{"string"}\NormalTok{)}
\NormalTok{            .}\BuiltInTok{str}\NormalTok{.replace(}\VerbatimStringTok{r"}\PreprocessorTok{[$,]}\VerbatimStringTok{"}\NormalTok{, }\StringTok{""}\NormalTok{, regex}\OperatorTok{=}\VariableTok{True}\NormalTok{)}
\NormalTok{            .astype(}\StringTok{"Float64"}\NormalTok{)}
\NormalTok{        ),}
\NormalTok{        margin\_pct}\OperatorTok{=}\KeywordTok{lambda}\NormalTok{ d: (}
\NormalTok{            d[}\StringTok{"margin\_pct"}\NormalTok{]}
\NormalTok{            .astype(}\StringTok{"string"}\NormalTok{)}
\NormalTok{            .}\BuiltInTok{str}\NormalTok{.rstrip(}\StringTok{"\%"}\NormalTok{)}
\NormalTok{            .astype(}\StringTok{"Float64"}\NormalTok{)}
\NormalTok{            .div(}\DecValTok{100}\NormalTok{)}
\NormalTok{        ),}
\NormalTok{        region}\OperatorTok{=}\KeywordTok{lambda}\NormalTok{ d: d[}\StringTok{"region"}\NormalTok{].astype(}\StringTok{"category"}\NormalTok{),}
\NormalTok{    )}
\NormalTok{)}
\end{Highlighting}
\end{Shaded}

This example captures a common business situation: a workbook contains
subtotal rows, blank rows, currency symbols, percent signs, inconsistent
date formats, and column names with extra spaces. A spreadsheet user can
fix these manually. \texttt{pandas} makes the rules explicit so the same
cleanup can run each month. Table 1 (Section 2) and the dtype practices
below address the cleaning and ingestion categories in Table 3.

\textbf{Supplementary Table 1.} Common Excel display issues and target
pandas dtypes

{\def\LTcaptype{none} 
\begin{longtable}[]{@{}
  >{\raggedright\arraybackslash}p{(\linewidth - 4\tabcolsep) * \real{0.3333}}
  >{\raggedright\arraybackslash}p{(\linewidth - 4\tabcolsep) * \real{0.3333}}
  >{\raggedright\arraybackslash}p{(\linewidth - 4\tabcolsep) * \real{0.3333}}@{}}
\toprule\noalign{}
\begin{minipage}[b]{\linewidth}\raggedright
Column as seen in Excel
\end{minipage} & \begin{minipage}[b]{\linewidth}\raggedright
Problem if ignored
\end{minipage} & \begin{minipage}[b]{\linewidth}\raggedright
Target pandas treatment
\end{minipage} \\
\midrule\noalign{}
\endhead
\bottomrule\noalign{}
\endlastfoot
\texttt{00421} (numeric ID) & Leading zeros lost &
\texttt{dtype="string"} at read \\
\texttt{\$1,250.50} & Arithmetic fails & Strip symbols -\textgreater{}
\texttt{Float64} \\
\texttt{32\%} & Treated as text or 32 & Parse as a 0-to-1 proportion \\
\texttt{1/5/2025} vs \texttt{2025-01-05} & Ambiguous sort order &
\texttt{pd.to\_datetime(...,\ errors="coerce")} \\
\texttt{Total} row in detail sheet & Double-count in pivots & Filter
with boolean mask \\
Blank cell vs empty string & Distinct missing semantics &
\texttt{.replace(r"\^{}\textbackslash{}s*\$",\ pd.NA)} \\
\end{longtable}
}

\subsection{4.2 Reading Excel Workbooks}\label{reading-excel-workbooks}

Many \texttt{pandas} workflows begin with Excel files. The goal should
be to read the workbook as data, not to preserve its visual layout.
Merged headers, blank spacer rows, subtotals, and human-oriented
formatting should be treated as ingestion problems.

\begin{Shaded}
\begin{Highlighting}[]
\ImportTok{from}\NormalTok{ pathlib }\ImportTok{import}\NormalTok{ Path}
\ImportTok{import}\NormalTok{ pandas }\ImportTok{as}\NormalTok{ pd}

\NormalTok{workbook }\OperatorTok{=}\NormalTok{ Path(}\StringTok{"monthly\_sales\_extract.xlsx"}\NormalTok{)}

\NormalTok{sales }\OperatorTok{=}\NormalTok{ pd.read\_excel(}
\NormalTok{    workbook,}
\NormalTok{    sheet\_name}\OperatorTok{=}\StringTok{"Sales"}\NormalTok{,}
\NormalTok{    dtype}\OperatorTok{=}\NormalTok{\{}
        \StringTok{"order\_id"}\NormalTok{: }\StringTok{"string"}\NormalTok{,}
        \StringTok{"customer\_id"}\NormalTok{: }\StringTok{"string"}\NormalTok{,}
        \StringTok{"product\_id"}\NormalTok{: }\StringTok{"string"}\NormalTok{,}
\NormalTok{    \},}
\NormalTok{    parse\_dates}\OperatorTok{=}\NormalTok{[}\StringTok{"order\_date"}\NormalTok{],}
\NormalTok{)}

\NormalTok{customers }\OperatorTok{=}\NormalTok{ pd.read\_excel(}
\NormalTok{    workbook,}
\NormalTok{    sheet\_name}\OperatorTok{=}\StringTok{"Customers"}\NormalTok{,}
\NormalTok{    dtype}\OperatorTok{=}\NormalTok{\{}\StringTok{"customer\_id"}\NormalTok{: }\StringTok{"string"}\NormalTok{\},}
\NormalTok{)}

\NormalTok{products }\OperatorTok{=}\NormalTok{ pd.read\_excel(}
\NormalTok{    workbook,}
\NormalTok{    sheet\_name}\OperatorTok{=}\StringTok{"Products"}\NormalTok{,}
\NormalTok{    dtype}\OperatorTok{=}\NormalTok{\{}\StringTok{"product\_id"}\NormalTok{: }\StringTok{"string"}\NormalTok{\},}
\NormalTok{)}
\end{Highlighting}
\end{Shaded}

When importing from Excel, analysts should prefer stable table ranges or
exported tabs over visually complex worksheets. If the workbook is also
used for presentation, it is often better to maintain a raw data sheet
separately from formatted output sheets.

\section{5. Replacing Spreadsheet Formulas with Table
Operations}\label{replacing-spreadsheet-formulas-with-table-operations}

The spreadsheet functions most often replaced in business analytics are
lookup formulas, pivot tables, filtering, conditional calculations, and
manual report formatting. The examples below show \texttt{pandas}
equivalents with validation.

\subsection{5.1 VLOOKUP and XLOOKUP as Validated
Joins}\label{vlookup-and-xlookup-as-validated-joins}

The direct analog to VLOOKUP or XLOOKUP is \texttt{merge()}. The
difference is that \texttt{pandas} can validate the expected
relationship between tables. This is important because accidental
duplicate keys in a lookup table can multiply rows and inflate totals.

\begin{Shaded}
\begin{Highlighting}[]
\ImportTok{import}\NormalTok{ pandas }\ImportTok{as}\NormalTok{ pd}

\NormalTok{orders }\OperatorTok{=}\NormalTok{ pd.DataFrame(\{}
    \StringTok{"order\_id"}\NormalTok{: [}\DecValTok{1001}\NormalTok{, }\DecValTok{1002}\NormalTok{, }\DecValTok{1003}\NormalTok{, }\DecValTok{1004}\NormalTok{, }\DecValTok{1005}\NormalTok{],}
    \StringTok{"customer\_id"}\NormalTok{: [}\StringTok{"C001"}\NormalTok{, }\StringTok{"C002"}\NormalTok{, }\StringTok{"C001"}\NormalTok{, }\StringTok{"C003"}\NormalTok{, }\StringTok{"C004"}\NormalTok{],}
    \StringTok{"product\_id"}\NormalTok{: [}\StringTok{"P10"}\NormalTok{, }\StringTok{"P20"}\NormalTok{, }\StringTok{"P10"}\NormalTok{, }\StringTok{"P30"}\NormalTok{, }\StringTok{"P20"}\NormalTok{],}
    \StringTok{"order\_date"}\NormalTok{: pd.to\_datetime([}
        \StringTok{"2025{-}01{-}05"}\NormalTok{, }\StringTok{"2025{-}01{-}07"}\NormalTok{, }\StringTok{"2025{-}02{-}03"}\NormalTok{,}
        \StringTok{"2025{-}02{-}20"}\NormalTok{, }\StringTok{"2025{-}02{-}25"}\NormalTok{,}
\NormalTok{    ]),}
    \StringTok{"quantity"}\NormalTok{: [}\DecValTok{3}\NormalTok{, }\DecValTok{5}\NormalTok{, }\DecValTok{2}\NormalTok{, }\DecValTok{8}\NormalTok{, }\DecValTok{4}\NormalTok{],}
    \StringTok{"unit\_price"}\NormalTok{: [}\FloatTok{120.00}\NormalTok{, }\FloatTok{45.50}\NormalTok{, }\FloatTok{120.00}\NormalTok{, }\FloatTok{15.25}\NormalTok{, }\FloatTok{45.50}\NormalTok{],}
\NormalTok{\})}

\NormalTok{customers }\OperatorTok{=}\NormalTok{ pd.DataFrame(\{}
    \StringTok{"customer\_id"}\NormalTok{: [}\StringTok{"C001"}\NormalTok{, }\StringTok{"C002"}\NormalTok{, }\StringTok{"C003"}\NormalTok{, }\StringTok{"C004"}\NormalTok{],}
    \StringTok{"customer\_name"}\NormalTok{: [}\StringTok{"Acme Inc."}\NormalTok{, }\StringTok{"Northwind"}\NormalTok{, }\StringTok{"Prairie Health"}\NormalTok{, }\StringTok{"Blue River"}\NormalTok{],}
    \StringTok{"region"}\NormalTok{: [}\StringTok{"West"}\NormalTok{, }\StringTok{"East"}\NormalTok{, }\StringTok{"Central"}\NormalTok{, }\StringTok{"West"}\NormalTok{],}
    \StringTok{"segment"}\NormalTok{: [}\StringTok{"Enterprise"}\NormalTok{, }\StringTok{"SMB"}\NormalTok{, }\StringTok{"Public"}\NormalTok{, }\StringTok{"SMB"}\NormalTok{],}
\NormalTok{\})}

\NormalTok{products }\OperatorTok{=}\NormalTok{ pd.DataFrame(\{}
    \StringTok{"product\_id"}\NormalTok{: [}\StringTok{"P10"}\NormalTok{, }\StringTok{"P20"}\NormalTok{, }\StringTok{"P30"}\NormalTok{],}
    \StringTok{"category"}\NormalTok{: [}\StringTok{"Hardware"}\NormalTok{, }\StringTok{"Subscription"}\NormalTok{, }\StringTok{"Services"}\NormalTok{],}
    \StringTok{"margin\_rate"}\NormalTok{: [}\FloatTok{0.31}\NormalTok{, }\FloatTok{0.74}\NormalTok{, }\FloatTok{0.46}\NormalTok{],}
\NormalTok{\})}

\NormalTok{enriched }\OperatorTok{=}\NormalTok{ (}
\NormalTok{    orders}
\NormalTok{    .merge(}
\NormalTok{        customers,}
\NormalTok{        on}\OperatorTok{=}\StringTok{"customer\_id"}\NormalTok{,}
\NormalTok{        how}\OperatorTok{=}\StringTok{"left"}\NormalTok{,}
\NormalTok{        validate}\OperatorTok{=}\StringTok{"m:1"}\NormalTok{,}
\NormalTok{        indicator}\OperatorTok{=}\StringTok{"customer\_match"}\NormalTok{,}
\NormalTok{    )}
\NormalTok{    .merge(}
\NormalTok{        products,}
\NormalTok{        on}\OperatorTok{=}\StringTok{"product\_id"}\NormalTok{,}
\NormalTok{        how}\OperatorTok{=}\StringTok{"left"}\NormalTok{,}
\NormalTok{        validate}\OperatorTok{=}\StringTok{"m:1"}\NormalTok{,}
\NormalTok{    )}
\NormalTok{)}

\NormalTok{missing\_customers }\OperatorTok{=}\NormalTok{ enriched.loc[}
\NormalTok{    enriched[}\StringTok{"customer\_match"}\NormalTok{].ne(}\StringTok{"both"}\NormalTok{),}
\NormalTok{    [}\StringTok{"order\_id"}\NormalTok{, }\StringTok{"customer\_id"}\NormalTok{],}
\NormalTok{]}
\ControlFlowTok{assert}\NormalTok{ missing\_customers.empty, }\SpecialStringTok{f"Unknown customer ids: }\SpecialCharTok{\{}\NormalTok{missing\_customers}\SpecialCharTok{.}\NormalTok{to\_dict(}\StringTok{\textquotesingle{}records\textquotesingle{}}\NormalTok{)}\SpecialCharTok{\}}\SpecialStringTok{"}

\NormalTok{enriched }\OperatorTok{=}\NormalTok{ (}
\NormalTok{    enriched}
\NormalTok{    .drop(columns}\OperatorTok{=}\StringTok{"customer\_match"}\NormalTok{)}
\NormalTok{    .assign(}
\NormalTok{        revenue}\OperatorTok{=}\KeywordTok{lambda}\NormalTok{ d: d[}\StringTok{"quantity"}\NormalTok{] }\OperatorTok{*}\NormalTok{ d[}\StringTok{"unit\_price"}\NormalTok{],}
\NormalTok{        gross\_margin}\OperatorTok{=}\KeywordTok{lambda}\NormalTok{ d: d[}\StringTok{"revenue"}\NormalTok{] }\OperatorTok{*}\NormalTok{ d[}\StringTok{"margin\_rate"}\NormalTok{],}
\NormalTok{    )}
\NormalTok{)}
\end{Highlighting}
\end{Shaded}

The \texttt{validate="m:1"} argument states that many orders may match
one customer or one product. If the customer table contains duplicate
\texttt{customer\_id} values, the join fails rather than quietly
producing duplicated revenue. This kind of guardrail is difficult to
enforce consistently in spreadsheets.

\subsection{\texorpdfstring{5.2 Pivot Tables as \texttt{groupby()} and
\texttt{pivot\_table()}}{5.2 Pivot Tables as groupby() and pivot\_table()}}\label{pivot-tables-as-groupby-and-pivot_table}

Pivot tables are one of the strongest bridges between Excel and
\texttt{pandas}. The analytical idea is the same: group by one or more
dimensions, aggregate measures, and reshape the result for reporting.

\begin{Shaded}
\begin{Highlighting}[]
\NormalTok{monthly\_metrics }\OperatorTok{=}\NormalTok{ (}
\NormalTok{    enriched}
\NormalTok{    .assign(month}\OperatorTok{=}\KeywordTok{lambda}\NormalTok{ d: d[}\StringTok{"order\_date"}\NormalTok{].dt.to\_period(}\StringTok{"M"}\NormalTok{))}
\NormalTok{    .groupby([}\StringTok{"month"}\NormalTok{, }\StringTok{"region"}\NormalTok{, }\StringTok{"segment"}\NormalTok{], observed}\OperatorTok{=}\VariableTok{True}\NormalTok{)}
\NormalTok{    .agg(}
\NormalTok{        revenue}\OperatorTok{=}\NormalTok{(}\StringTok{"revenue"}\NormalTok{, }\StringTok{"sum"}\NormalTok{),}
\NormalTok{        gross\_margin}\OperatorTok{=}\NormalTok{(}\StringTok{"gross\_margin"}\NormalTok{, }\StringTok{"sum"}\NormalTok{),}
\NormalTok{        orders}\OperatorTok{=}\NormalTok{(}\StringTok{"order\_id"}\NormalTok{, }\StringTok{"nunique"}\NormalTok{),}
\NormalTok{        units}\OperatorTok{=}\NormalTok{(}\StringTok{"quantity"}\NormalTok{, }\StringTok{"sum"}\NormalTok{),}
\NormalTok{    )}
\NormalTok{    .assign(}
\NormalTok{        average\_order\_value}\OperatorTok{=}\KeywordTok{lambda}\NormalTok{ d: d[}\StringTok{"revenue"}\NormalTok{] }\OperatorTok{/}\NormalTok{ d[}\StringTok{"orders"}\NormalTok{],}
\NormalTok{        margin\_rate}\OperatorTok{=}\KeywordTok{lambda}\NormalTok{ d: d[}\StringTok{"gross\_margin"}\NormalTok{] }\OperatorTok{/}\NormalTok{ d[}\StringTok{"revenue"}\NormalTok{],}
\NormalTok{    )}
\NormalTok{    .reset\_index()}
\NormalTok{    .sort\_values([}\StringTok{"month"}\NormalTok{, }\StringTok{"region"}\NormalTok{, }\StringTok{"segment"}\NormalTok{])}
\NormalTok{)}

\NormalTok{regional\_dashboard }\OperatorTok{=}\NormalTok{ (}
\NormalTok{    monthly\_metrics}
\NormalTok{    .pivot\_table(}
\NormalTok{        index}\OperatorTok{=}\StringTok{"month"}\NormalTok{,}
\NormalTok{        columns}\OperatorTok{=}\StringTok{"region"}\NormalTok{,}
\NormalTok{        values}\OperatorTok{=}\StringTok{"revenue"}\NormalTok{,}
\NormalTok{        aggfunc}\OperatorTok{=}\StringTok{"sum"}\NormalTok{,}
\NormalTok{        fill\_value}\OperatorTok{=}\DecValTok{0}\NormalTok{,}
\NormalTok{    )}
\NormalTok{    .sort\_index()}
\NormalTok{)}
\end{Highlighting}
\end{Shaded}

In Excel, the pivot table object stores configuration inside a workbook.
In \texttt{pandas}, the grouping logic is visible in the script. This
makes the transformation easier to review, rerun, and compare across
versions.

\subsection{5.3 Business Rule
Validation}\label{business-rule-validation}

Business analytics is not only about computing metrics. It is also about
proving that the data satisfies assumptions before numbers are
distributed. Assertions and reconciliation tables should be normal parts
of \texttt{pandas} workflows.

\begin{Shaded}
\begin{Highlighting}[]
\ImportTok{import}\NormalTok{ numpy }\ImportTok{as}\NormalTok{ np}

\NormalTok{required\_columns }\OperatorTok{=}\NormalTok{ \{}
    \StringTok{"order\_id"}\NormalTok{, }\StringTok{"customer\_id"}\NormalTok{, }\StringTok{"product\_id"}\NormalTok{, }\StringTok{"order\_date"}\NormalTok{,}
    \StringTok{"quantity"}\NormalTok{, }\StringTok{"unit\_price"}\NormalTok{, }\StringTok{"revenue"}\NormalTok{,}
\NormalTok{\}}

\NormalTok{missing\_columns }\OperatorTok{=}\NormalTok{ required\_columns.difference(enriched.columns)}
\ControlFlowTok{assert} \KeywordTok{not}\NormalTok{ missing\_columns, }\SpecialStringTok{f"Missing required columns: }\SpecialCharTok{\{}\BuiltInTok{sorted}\NormalTok{(missing\_columns)}\SpecialCharTok{\}}\SpecialStringTok{"}

\ControlFlowTok{assert}\NormalTok{ enriched[}\StringTok{"order\_id"}\NormalTok{].is\_unique, }\StringTok{"Order identifiers should be unique in this extract."}
\ControlFlowTok{assert}\NormalTok{ enriched[}\StringTok{"order\_date"}\NormalTok{].notna().}\BuiltInTok{all}\NormalTok{(), }\StringTok{"Every order must have a valid order date."}
\ControlFlowTok{assert}\NormalTok{ (enriched[}\StringTok{"quantity"}\NormalTok{] }\OperatorTok{\textgreater{}} \DecValTok{0}\NormalTok{).}\BuiltInTok{all}\NormalTok{(), }\StringTok{"Quantities must be positive."}

\NormalTok{revenue\_from\_detail }\OperatorTok{=}\NormalTok{ enriched[}\StringTok{"revenue"}\NormalTok{].}\BuiltInTok{sum}\NormalTok{()}
\NormalTok{revenue\_from\_report }\OperatorTok{=}\NormalTok{ monthly\_metrics[}\StringTok{"revenue"}\NormalTok{].}\BuiltInTok{sum}\NormalTok{()}
\ControlFlowTok{assert}\NormalTok{ np.isclose(}
\NormalTok{    revenue\_from\_detail,}
\NormalTok{    revenue\_from\_report,}
\NormalTok{), }\StringTok{"Report does not reconcile to detail."}

\NormalTok{quality\_report }\OperatorTok{=}\NormalTok{ pd.DataFrame(\{}
    \StringTok{"check"}\NormalTok{: [}\StringTok{"orders"}\NormalTok{, }\StringTok{"customers"}\NormalTok{, }\StringTok{"products"}\NormalTok{, }\StringTok{"revenue"}\NormalTok{, }\StringTok{"gross\_margin"}\NormalTok{],}
    \StringTok{"value"}\NormalTok{: [}
\NormalTok{        enriched[}\StringTok{"order\_id"}\NormalTok{].nunique(),}
\NormalTok{        enriched[}\StringTok{"customer\_id"}\NormalTok{].nunique(),}
\NormalTok{        enriched[}\StringTok{"product\_id"}\NormalTok{].nunique(),}
\NormalTok{        enriched[}\StringTok{"revenue"}\NormalTok{].}\BuiltInTok{sum}\NormalTok{(),}
\NormalTok{        enriched[}\StringTok{"gross\_margin"}\NormalTok{].}\BuiltInTok{sum}\NormalTok{(),}
\NormalTok{    ],}
\NormalTok{\})}
\end{Highlighting}
\end{Shaded}

These checks are simple. Their value comes from being automatic. A
workbook can look correct while containing a duplicate lookup key, a
missing product, a text-formatted date, or a stale pasted value. A
script can refuse to produce a report until the assumptions are
satisfied. The \texttt{quality\_report} object is a compact audit
artifact suitable for attaching to monthly close packages:

{\def\LTcaptype{none} 
\begin{longtable}[]{@{}ll@{}}
\toprule\noalign{}
check & value (example) \\
\midrule\noalign{}
\endhead
\bottomrule\noalign{}
\endlastfoot
orders & 5 \\
customers & 4 \\
products & 3 \\
revenue & 2,847.00 \\
gross\_margin & 1,456.23 \\
\end{longtable}
}

\subsection{5.4 Exporting Stakeholder-Friendly Excel
Reports}\label{exporting-stakeholder-friendly-excel-reports}

Many organizations still need Excel output. This is compatible with a
reproducible workflow. \texttt{pandas} can generate workbook tabs after
the calculations are complete.

\begin{Shaded}
\begin{Highlighting}[]
\ControlFlowTok{with}\NormalTok{ pd.ExcelWriter(}
    \StringTok{"monthly\_business\_report.xlsx"}\NormalTok{,}
\NormalTok{    engine}\OperatorTok{=}\StringTok{"xlsxwriter"}\NormalTok{,}
\NormalTok{    datetime\_format}\OperatorTok{=}\StringTok{"yyyy{-}mm{-}dd"}\NormalTok{,}
\NormalTok{    date\_format}\OperatorTok{=}\StringTok{"yyyy{-}mm{-}dd"}\NormalTok{,}
\NormalTok{) }\ImportTok{as}\NormalTok{ writer:}
\NormalTok{    monthly\_metrics.to\_excel(writer, sheet\_name}\OperatorTok{=}\StringTok{"Monthly metrics"}\NormalTok{, index}\OperatorTok{=}\VariableTok{False}\NormalTok{)}
\NormalTok{    regional\_dashboard.to\_excel(writer, sheet\_name}\OperatorTok{=}\StringTok{"Regional dashboard"}\NormalTok{)}
\NormalTok{    quality\_report.to\_excel(writer, sheet\_name}\OperatorTok{=}\StringTok{"Data quality"}\NormalTok{, index}\OperatorTok{=}\VariableTok{False}\NormalTok{)}

\NormalTok{    workbook }\OperatorTok{=}\NormalTok{ writer.book}
\NormalTok{    money\_format }\OperatorTok{=}\NormalTok{ workbook.add\_format(\{}\StringTok{"num\_format"}\NormalTok{: }\StringTok{"$\#,\#\#0"}\NormalTok{\})}
\NormalTok{    metrics\_sheet }\OperatorTok{=}\NormalTok{ writer.sheets[}\StringTok{"Monthly metrics"}\NormalTok{]}
\NormalTok{    metrics\_sheet.freeze\_panes(}\DecValTok{1}\NormalTok{, }\DecValTok{0}\NormalTok{)}
\NormalTok{    metrics\_sheet.set\_column(}\StringTok{"D:E"}\NormalTok{, }\DecValTok{14}\NormalTok{, money\_format)}
\end{Highlighting}
\end{Shaded}

This pattern separates computation from presentation. The workbook
becomes an output artifact, not the only place where business logic
exists.

\section{6. End-to-End Applied
Examples}\label{end-to-end-applied-examples}

The following examples illustrate full workflows that combine ingestion,
cleaning, validation, aggregation, and reporting. They use sample
business and research datasets. Table 4 maps each example to the
workflow patterns in Table 3; Figures 4--6 visualize representative
pipelines and outputs.

\textbf{Table 4.} End-to-end examples and workflow patterns illustrated

{\def\LTcaptype{none} 
\begin{longtable}[]{@{}
  >{\raggedright\arraybackslash}p{(\linewidth - 6\tabcolsep) * \real{0.2500}}
  >{\raggedright\arraybackslash}p{(\linewidth - 6\tabcolsep) * \real{0.2500}}
  >{\raggedright\arraybackslash}p{(\linewidth - 6\tabcolsep) * \real{0.2500}}
  >{\raggedright\arraybackslash}p{(\linewidth - 6\tabcolsep) * \real{0.2500}}@{}}
\toprule\noalign{}
\begin{minipage}[b]{\linewidth}\raggedright
Example
\end{minipage} & \begin{minipage}[b]{\linewidth}\raggedright
Domain
\end{minipage} & \begin{minipage}[b]{\linewidth}\raggedright
Patterns emphasized (Table 3)
\end{minipage} & \begin{minipage}[b]{\linewidth}\raggedright
Primary outputs
\end{minipage} \\
\midrule\noalign{}
\endhead
\bottomrule\noalign{}
\endlastfoot
6.1 Monthly sales reconciliation & Business analytics & 1--3, 5--7, 9 &
Detail + summary Excel, reconciliation \\
6.2 Messy Excel intake form & Operations / admin & 1--2, 6 & Clean
submission table \\
6.3 Survey data workflow & Applied research & 1--2, 4--6 & Long-format
data, dept summaries \\
6.4 Customer complaint analysis & Service operations & 1--2, 5--7 &
Management summary, overdue flags \\
6.5 SQL-to-pandas workflow & Hybrid architecture & 1, 5--6, 9 &
Validated aggregate table in DB \\
6.6 Time series monitoring & Operations / IoT & 1, 5--7 & Daily rollups,
anomaly flags \\
6.7 Machine learning preparation & Predictive analytics & 5, 8, 9 &
sklearn pipeline, CV scores \\
\end{longtable}
}

\subsection{6.1 Example 1: Monthly Sales
Reconciliation}\label{example-1-monthly-sales-reconciliation}

Monthly sales close is a common business analytics task in which order
detail must agree with regional or product summaries before results are
distributed. This example reads three related workbook sheets---orders,
customers, and products---then enriches the order table through
validated left joins that reject duplicate lookup keys. Revenue and
gross margin are computed at the line level, aggregated by month and
region, and checked so that detail totals equal summary totals; only
then is a multi-sheet Excel report written. Figure 4 shows the pipeline
flow; Figure 5 shows a sample pivot-style dashboard output.

\textbf{Figure 4.} Monthly sales reconciliation pipeline flowchart
(Example 6.1).

\begin{center}
\includegraphics[width=0.96\linewidth,height=0.62\textheight,keepaspectratio]{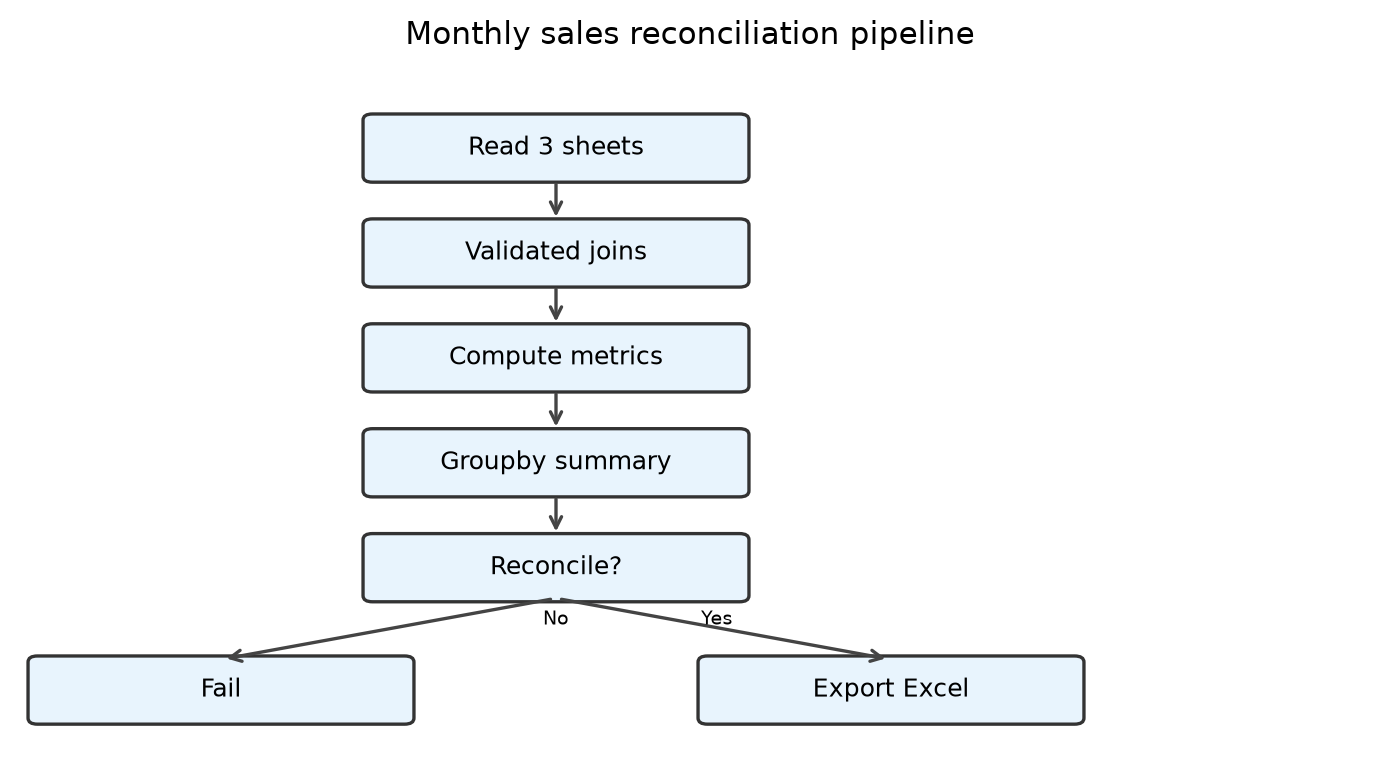}
\end{center}

\begin{Shaded}
\begin{Highlighting}[]
\ImportTok{import}\NormalTok{ numpy }\ImportTok{as}\NormalTok{ np}
\ImportTok{import}\NormalTok{ pandas }\ImportTok{as}\NormalTok{ pd}
\ImportTok{from}\NormalTok{ pathlib }\ImportTok{import}\NormalTok{ Path}

\KeywordTok{def}\NormalTok{ read\_sales\_workbook(path: Path) }\OperatorTok{{-}\textgreater{}} \BuiltInTok{tuple}\NormalTok{[pd.DataFrame, pd.DataFrame, pd.DataFrame]:}
\NormalTok{    orders }\OperatorTok{=}\NormalTok{ pd.read\_excel(}
\NormalTok{        path,}
\NormalTok{        sheet\_name}\OperatorTok{=}\StringTok{"Orders"}\NormalTok{,}
\NormalTok{        dtype}\OperatorTok{=}\NormalTok{\{}
            \StringTok{"order\_id"}\NormalTok{: }\StringTok{"string"}\NormalTok{,}
            \StringTok{"customer\_id"}\NormalTok{: }\StringTok{"string"}\NormalTok{,}
            \StringTok{"product\_id"}\NormalTok{: }\StringTok{"string"}\NormalTok{,}
\NormalTok{        \},}
\NormalTok{        parse\_dates}\OperatorTok{=}\NormalTok{[}\StringTok{"order\_date"}\NormalTok{],}
\NormalTok{    )}
\NormalTok{    customers }\OperatorTok{=}\NormalTok{ pd.read\_excel(path, sheet\_name}\OperatorTok{=}\StringTok{"Customers"}\NormalTok{, dtype}\OperatorTok{=}\NormalTok{\{}\StringTok{"customer\_id"}\NormalTok{: }\StringTok{"string"}\NormalTok{\})}
\NormalTok{    products }\OperatorTok{=}\NormalTok{ pd.read\_excel(path, sheet\_name}\OperatorTok{=}\StringTok{"Products"}\NormalTok{, dtype}\OperatorTok{=}\NormalTok{\{}\StringTok{"product\_id"}\NormalTok{: }\StringTok{"string"}\NormalTok{\})}
    \ControlFlowTok{return}\NormalTok{ orders, customers, products}

\KeywordTok{def}\NormalTok{ build\_enriched\_sales(orders, customers, products) }\OperatorTok{{-}\textgreater{}}\NormalTok{ pd.DataFrame:}
\NormalTok{    detail }\OperatorTok{=}\NormalTok{ (}
\NormalTok{        orders}
\NormalTok{        .merge(customers, on}\OperatorTok{=}\StringTok{"customer\_id"}\NormalTok{, how}\OperatorTok{=}\StringTok{"left"}\NormalTok{, validate}\OperatorTok{=}\StringTok{"m:1"}\NormalTok{, indicator}\OperatorTok{=}\StringTok{"cust\_ok"}\NormalTok{)}
\NormalTok{        .merge(products, on}\OperatorTok{=}\StringTok{"product\_id"}\NormalTok{, how}\OperatorTok{=}\StringTok{"left"}\NormalTok{, validate}\OperatorTok{=}\StringTok{"m:1"}\NormalTok{, indicator}\OperatorTok{=}\StringTok{"prod\_ok"}\NormalTok{)}
\NormalTok{    )}
\NormalTok{    bad }\OperatorTok{=}\NormalTok{ detail.loc[detail[}\StringTok{"cust\_ok"}\NormalTok{].ne(}\StringTok{"both"}\NormalTok{) }\OperatorTok{|}\NormalTok{ detail[}\StringTok{"prod\_ok"}\NormalTok{].ne(}\StringTok{"both"}\NormalTok{)]}
\NormalTok{    key\_cols }\OperatorTok{=}\NormalTok{ [}\StringTok{"order\_id"}\NormalTok{, }\StringTok{"customer\_id"}\NormalTok{, }\StringTok{"product\_id"}\NormalTok{]}
    \ControlFlowTok{assert}\NormalTok{ bad.empty, }\SpecialStringTok{f"Unmatched keys: }\SpecialCharTok{\{}\NormalTok{bad[key\_cols]}\SpecialCharTok{.}\NormalTok{head()}\SpecialCharTok{.}\NormalTok{to\_dict(}\StringTok{\textquotesingle{}records\textquotesingle{}}\NormalTok{)}\SpecialCharTok{\}}\SpecialStringTok{"}
    \ControlFlowTok{return}\NormalTok{ detail.drop(columns}\OperatorTok{=}\NormalTok{[}\StringTok{"cust\_ok"}\NormalTok{, }\StringTok{"prod\_ok"}\NormalTok{]).assign(}
\NormalTok{        revenue}\OperatorTok{=}\KeywordTok{lambda}\NormalTok{ d: d[}\StringTok{"quantity"}\NormalTok{] }\OperatorTok{*}\NormalTok{ d[}\StringTok{"unit\_price"}\NormalTok{],}
\NormalTok{        gross\_margin}\OperatorTok{=}\KeywordTok{lambda}\NormalTok{ d: d[}\StringTok{"revenue"}\NormalTok{] }\OperatorTok{*}\NormalTok{ d[}\StringTok{"margin\_rate"}\NormalTok{],}
\NormalTok{        month}\OperatorTok{=}\KeywordTok{lambda}\NormalTok{ d: d[}\StringTok{"order\_date"}\NormalTok{].dt.to\_period(}\StringTok{"M"}\NormalTok{),}
\NormalTok{    )}

\NormalTok{path }\OperatorTok{=}\NormalTok{ Path(}\StringTok{"data/raw/monthly\_sales.xlsx"}\NormalTok{)}
\NormalTok{orders, customers, products }\OperatorTok{=}\NormalTok{ read\_sales\_workbook(path)}
\NormalTok{enriched }\OperatorTok{=}\NormalTok{ build\_enriched\_sales(orders, customers, products)}

\NormalTok{summary }\OperatorTok{=}\NormalTok{ enriched.groupby([}\StringTok{"month"}\NormalTok{, }\StringTok{"region"}\NormalTok{], observed}\OperatorTok{=}\VariableTok{True}\NormalTok{).agg(}
\NormalTok{    revenue}\OperatorTok{=}\NormalTok{(}\StringTok{"revenue"}\NormalTok{, }\StringTok{"sum"}\NormalTok{),}
\NormalTok{    gross\_margin}\OperatorTok{=}\NormalTok{(}\StringTok{"gross\_margin"}\NormalTok{, }\StringTok{"sum"}\NormalTok{),}
\NormalTok{    orders}\OperatorTok{=}\NormalTok{(}\StringTok{"order\_id"}\NormalTok{, }\StringTok{"nunique"}\NormalTok{),}
\NormalTok{).reset\_index()}

\ControlFlowTok{assert}\NormalTok{ np.isclose(enriched[}\StringTok{"revenue"}\NormalTok{].}\BuiltInTok{sum}\NormalTok{(), summary[}\StringTok{"revenue"}\NormalTok{].}\BuiltInTok{sum}\NormalTok{())}

\ControlFlowTok{with}\NormalTok{ pd.ExcelWriter(}
    \StringTok{"reports/monthly\_sales\_reconciliation.xlsx"}\NormalTok{,}
\NormalTok{    engine}\OperatorTok{=}\StringTok{"xlsxwriter"}\NormalTok{,}
\NormalTok{) }\ImportTok{as}\NormalTok{ writer:}
\NormalTok{    enriched.to\_excel(writer, sheet\_name}\OperatorTok{=}\StringTok{"Detail"}\NormalTok{, index}\OperatorTok{=}\VariableTok{False}\NormalTok{)}
\NormalTok{    summary.to\_excel(writer, sheet\_name}\OperatorTok{=}\StringTok{"Summary"}\NormalTok{, index}\OperatorTok{=}\VariableTok{False}\NormalTok{)}
\end{Highlighting}
\end{Shaded}

\textbf{Figure 5.} Regional revenue dashboard produced by
\texttt{pivot\_table()} (Example 6.1).

\begin{center}
\includegraphics[width=0.96\linewidth,height=0.62\textheight,keepaspectratio]{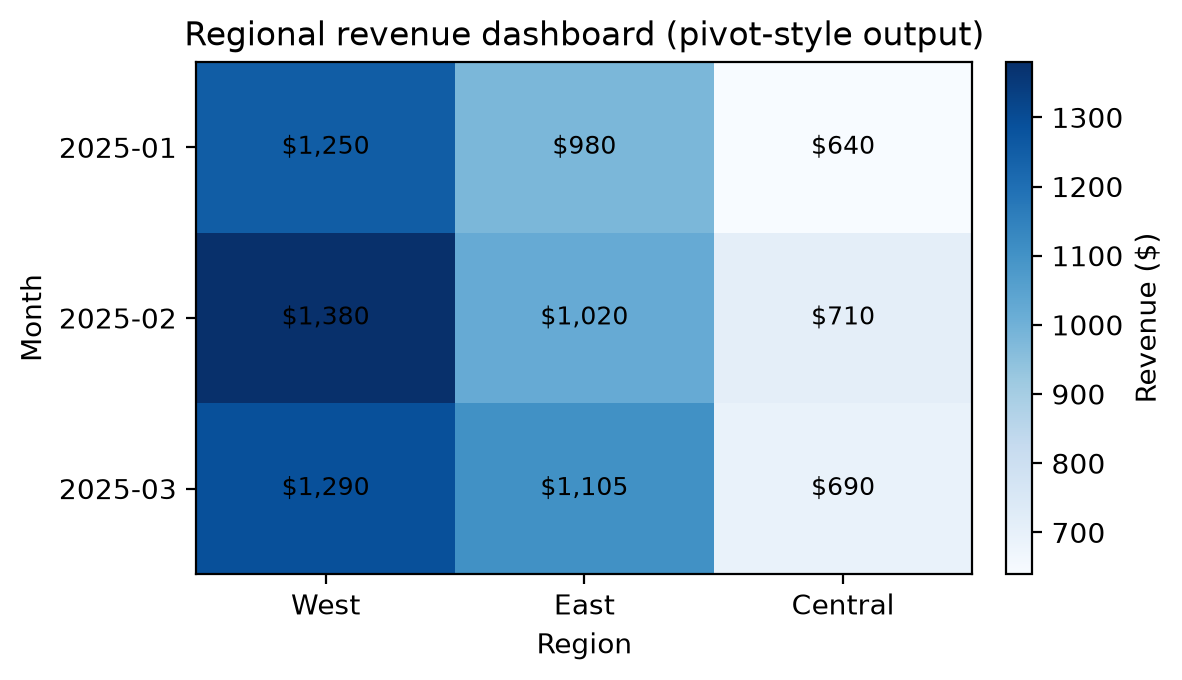}
\end{center}

\subsection{6.2 Example 2: Messy Excel Intake
Form}\label{example-2-messy-excel-intake-form}

Operational teams often receive Excel intake forms that were designed
for human reading rather than programmatic import. Title rows, blank
lines, embedded subtotals, and currency symbols can break a naive
\texttt{read\_excel()} call. This workflow locates the true header row,
standardizes column names, removes blank and subtotal rows, parses dates
and currency fields, and asserts that required columns, unique
submission IDs, and valid dates are all present before the table is used
downstream.

\begin{Shaded}
\begin{Highlighting}[]
\ImportTok{import}\NormalTok{ pandas }\ImportTok{as}\NormalTok{ pd}

\NormalTok{raw }\OperatorTok{=}\NormalTok{ pd.read\_excel(}\StringTok{"intake\_form.xlsx"}\NormalTok{, sheet\_name}\OperatorTok{=}\StringTok{"Submissions"}\NormalTok{, header}\OperatorTok{=}\VariableTok{None}\NormalTok{)}
\NormalTok{header\_row }\OperatorTok{=}\NormalTok{ raw.index[}
\NormalTok{    raw.iloc[:, }\DecValTok{0}\NormalTok{].astype(}\StringTok{"string"}\NormalTok{).}\BuiltInTok{str}\NormalTok{.contains(}\StringTok{"Submission ID"}\NormalTok{, na}\OperatorTok{=}\VariableTok{False}\NormalTok{)}
\NormalTok{][}\DecValTok{0}\NormalTok{]}
\NormalTok{data }\OperatorTok{=}\NormalTok{ pd.read\_excel(}\StringTok{"intake\_form.xlsx"}\NormalTok{, sheet\_name}\OperatorTok{=}\StringTok{"Submissions"}\NormalTok{, header}\OperatorTok{=}\NormalTok{header\_row)}

\NormalTok{clean }\OperatorTok{=}\NormalTok{ (}
\NormalTok{    data}
\NormalTok{    .rename(columns}\OperatorTok{=}\KeywordTok{lambda}\NormalTok{ c: }\BuiltInTok{str}\NormalTok{(c).strip().lower().replace(}\StringTok{" "}\NormalTok{, }\StringTok{"\_"}\NormalTok{))}
\NormalTok{    .dropna(how}\OperatorTok{=}\StringTok{"all"}\NormalTok{)}
\NormalTok{    .loc[}\KeywordTok{lambda}\NormalTok{ d: d[}\StringTok{"submission\_id"}\NormalTok{].notna()]}
\NormalTok{    .loc[}
        \KeywordTok{lambda}\NormalTok{ d: }\OperatorTok{\textasciitilde{}}\NormalTok{d[}\StringTok{"submission\_id"}\NormalTok{]}
\NormalTok{        .astype(}\StringTok{"string"}\NormalTok{)}
\NormalTok{        .}\BuiltInTok{str}\NormalTok{.lower()}
\NormalTok{        .isin([}\StringTok{"total"}\NormalTok{, }\StringTok{"subtotal"}\NormalTok{])}
\NormalTok{    ]}
\NormalTok{    .assign(}
\NormalTok{        submission\_id}\OperatorTok{=}\KeywordTok{lambda}\NormalTok{ d: d[}\StringTok{"submission\_id"}\NormalTok{].astype(}\StringTok{"string"}\NormalTok{),}
\NormalTok{        submitted\_on}\OperatorTok{=}\KeywordTok{lambda}\NormalTok{ d: pd.to\_datetime(d[}\StringTok{"submitted\_on"}\NormalTok{], errors}\OperatorTok{=}\StringTok{"coerce"}\NormalTok{),}
\NormalTok{        amount\_requested}\OperatorTok{=}\KeywordTok{lambda}\NormalTok{ d: (}
\NormalTok{            d[}\StringTok{"amount\_requested"}\NormalTok{]}
\NormalTok{            .astype(}\StringTok{"string"}\NormalTok{)}
\NormalTok{            .}\BuiltInTok{str}\NormalTok{.replace(}\VerbatimStringTok{r"}\PreprocessorTok{[$,]}\VerbatimStringTok{"}\NormalTok{, }\StringTok{""}\NormalTok{, regex}\OperatorTok{=}\VariableTok{True}\NormalTok{)}
\NormalTok{            .astype(}\StringTok{"Float64"}\NormalTok{)}
\NormalTok{        ),}
\NormalTok{    )}
\NormalTok{)}

\NormalTok{required }\OperatorTok{=}\NormalTok{ \{}
    \StringTok{"submission\_id"}\NormalTok{,}
    \StringTok{"submitted\_on"}\NormalTok{,}
    \StringTok{"department"}\NormalTok{,}
    \StringTok{"amount\_requested"}\NormalTok{,}
\NormalTok{\}}
\NormalTok{missing }\OperatorTok{=}\NormalTok{ required.difference(clean.columns)}
\ControlFlowTok{assert} \KeywordTok{not}\NormalTok{ missing, }\SpecialStringTok{f"Missing required columns: }\SpecialCharTok{\{}\BuiltInTok{sorted}\NormalTok{(missing)}\SpecialCharTok{\}}\SpecialStringTok{"}
\ControlFlowTok{assert}\NormalTok{ clean[}\StringTok{"submission\_id"}\NormalTok{].is\_unique}
\ControlFlowTok{assert}\NormalTok{ clean[}\StringTok{"submitted\_on"}\NormalTok{].notna().}\BuiltInTok{all}\NormalTok{()}
\end{Highlighting}
\end{Shaded}

\subsection{6.3 Example 3: Survey Data
Workflow}\label{example-3-survey-data-workflow}

Employee or customer survey tools typically export one row per
respondent with one column per question, using text labels rather than
numeric scores. That layout is easy to read in a spreadsheet but awkward
for aggregation and statistical modeling. This workflow maps Likert
labels to ordered categories, converts responses to numeric scores,
reshapes the table from wide to long format with \texttt{melt()},
computes department-level summaries, and records a small data-quality
table before further analysis. Figure 6 shows a department-level summary
after the wide-to-long reshape central to tidy analysis {[}6{]}.

\textbf{Figure 6.} Department-level survey summary after wide-to-long
reshaping (Example 6.3).

\begin{center}
\includegraphics[width=0.96\linewidth,height=0.62\textheight,keepaspectratio]{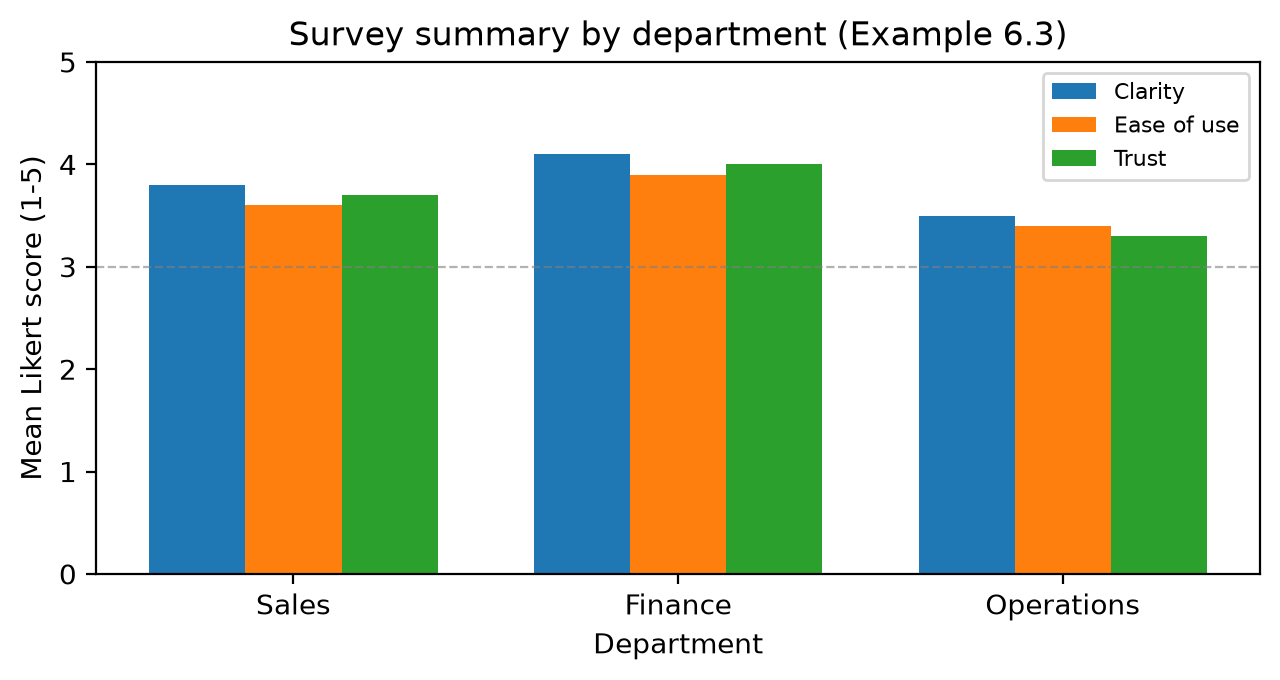}
\end{center}

\begin{Shaded}
\begin{Highlighting}[]
\ImportTok{import}\NormalTok{ pandas }\ImportTok{as}\NormalTok{ pd}

\NormalTok{survey }\OperatorTok{=}\NormalTok{ pd.read\_csv(}\StringTok{"data/raw/employee\_survey.csv"}\NormalTok{, dtype}\OperatorTok{=}\NormalTok{\{}\StringTok{"respondent\_id"}\NormalTok{: }\StringTok{"string"}\NormalTok{\})}
\NormalTok{likert\_order }\OperatorTok{=}\NormalTok{ [}\StringTok{"Strongly disagree"}\NormalTok{, }\StringTok{"Disagree"}\NormalTok{, }\StringTok{"Neutral"}\NormalTok{, }\StringTok{"Agree"}\NormalTok{, }\StringTok{"Strongly agree"}\NormalTok{]}
\NormalTok{likert\_score }\OperatorTok{=}\NormalTok{ \{label: i }\ControlFlowTok{for}\NormalTok{ i, label }\KeywordTok{in} \BuiltInTok{enumerate}\NormalTok{(likert\_order, start}\OperatorTok{=}\DecValTok{1}\NormalTok{)\}}
\NormalTok{questions }\OperatorTok{=}\NormalTok{ [}\StringTok{"clarity"}\NormalTok{, }\StringTok{"ease\_of\_use"}\NormalTok{, }\StringTok{"trust"}\NormalTok{]}

\NormalTok{clean }\OperatorTok{=}\NormalTok{ survey.assign(}
\NormalTok{    submitted\_at}\OperatorTok{=}\KeywordTok{lambda}\NormalTok{ d: pd.to\_datetime(d[}\StringTok{"submitted\_at"}\NormalTok{], errors}\OperatorTok{=}\StringTok{"coerce"}\NormalTok{)}
\NormalTok{)}
\ControlFlowTok{for}\NormalTok{ q }\KeywordTok{in}\NormalTok{ questions:}
\NormalTok{    clean[q] }\OperatorTok{=}\NormalTok{ pd.Categorical(clean[q], categories}\OperatorTok{=}\NormalTok{likert\_order, ordered}\OperatorTok{=}\VariableTok{True}\NormalTok{)}
\NormalTok{    clean[}\SpecialStringTok{f"}\SpecialCharTok{\{}\NormalTok{q}\SpecialCharTok{\}}\SpecialStringTok{\_score"}\NormalTok{] }\OperatorTok{=}\NormalTok{ clean[q].}\BuiltInTok{map}\NormalTok{(likert\_score).astype(}\StringTok{"Int64"}\NormalTok{)}

\NormalTok{survey\_long }\OperatorTok{=}\NormalTok{ clean.melt(}
\NormalTok{    id\_vars}\OperatorTok{=}\NormalTok{[}\StringTok{"respondent\_id"}\NormalTok{, }\StringTok{"department"}\NormalTok{, }\StringTok{"submitted\_at"}\NormalTok{],}
\NormalTok{    value\_vars}\OperatorTok{=}\NormalTok{[}\SpecialStringTok{f"}\SpecialCharTok{\{}\NormalTok{q}\SpecialCharTok{\}}\SpecialStringTok{\_score"} \ControlFlowTok{for}\NormalTok{ q }\KeywordTok{in}\NormalTok{ questions],}
\NormalTok{    var\_name}\OperatorTok{=}\StringTok{"question"}\NormalTok{,}
\NormalTok{    value\_name}\OperatorTok{=}\StringTok{"score"}\NormalTok{,}
\NormalTok{)}

\NormalTok{dept\_summary }\OperatorTok{=}\NormalTok{ survey\_long.groupby([}\StringTok{"department"}\NormalTok{, }\StringTok{"question"}\NormalTok{], observed}\OperatorTok{=}\VariableTok{True}\NormalTok{).agg(}
\NormalTok{    responses}\OperatorTok{=}\NormalTok{(}\StringTok{"score"}\NormalTok{, }\StringTok{"count"}\NormalTok{),}
\NormalTok{    mean\_score}\OperatorTok{=}\NormalTok{(}\StringTok{"score"}\NormalTok{, }\StringTok{"mean"}\NormalTok{),}
\NormalTok{    median\_score}\OperatorTok{=}\NormalTok{(}\StringTok{"score"}\NormalTok{, }\StringTok{"median"}\NormalTok{),}
\NormalTok{).reset\_index()}

\NormalTok{quality }\OperatorTok{=}\NormalTok{ pd.DataFrame(\{}
    \StringTok{"metric"}\NormalTok{: [}\StringTok{"respondents"}\NormalTok{, }\StringTok{"missing\_submitted\_at"}\NormalTok{, }\StringTok{"duplicate\_ids"}\NormalTok{],}
    \StringTok{"value"}\NormalTok{: [}
\NormalTok{        clean[}\StringTok{"respondent\_id"}\NormalTok{].nunique(),}
\NormalTok{        clean[}\StringTok{"submitted\_at"}\NormalTok{].isna().}\BuiltInTok{sum}\NormalTok{(),}
\NormalTok{        clean[}\StringTok{"respondent\_id"}\NormalTok{].duplicated().}\BuiltInTok{sum}\NormalTok{(),}
\NormalTok{    ],}
\NormalTok{\})}
\end{Highlighting}
\end{Shaded}

The long table \texttt{survey\_long} is ready for visualization, mixed
models, or export to R.

\subsection{6.4 Example 4: Customer Complaint
Analysis}\label{example-4-customer-complaint-analysis}

Customer complaint files combine timestamps, product categories,
severity codes, and channel labels that are often entered
inconsistently. Service managers need monthly summaries of case volume,
resolution time, and open or overdue work against stated SLAs. This
example standardizes channel names, derives resolution days and
open-month fields, flags overdue cases that remain unresolved past their
SLA, and aggregates results by month, product, and severity for a
management summary table.

\begin{Shaded}
\begin{Highlighting}[]
\ImportTok{import}\NormalTok{ pandas }\ImportTok{as}\NormalTok{ pd}

\NormalTok{complaints }\OperatorTok{=}\NormalTok{ pd.read\_excel(}
    \StringTok{"complaints.xlsx"}\NormalTok{,}
\NormalTok{    dtype}\OperatorTok{=}\NormalTok{\{}\StringTok{"case\_id"}\NormalTok{: }\StringTok{"string"}\NormalTok{, }\StringTok{"customer\_id"}\NormalTok{: }\StringTok{"string"}\NormalTok{\},}
\NormalTok{    parse\_dates}\OperatorTok{=}\NormalTok{[}\StringTok{"opened\_at"}\NormalTok{, }\StringTok{"closed\_at"}\NormalTok{],}
\NormalTok{)}

\NormalTok{channel\_map }\OperatorTok{=}\NormalTok{ \{}\StringTok{"phone"}\NormalTok{: }\StringTok{"Phone"}\NormalTok{, }\StringTok{"PHONE"}\NormalTok{: }\StringTok{"Phone"}\NormalTok{, }\StringTok{"web"}\NormalTok{: }\StringTok{"Web"}\NormalTok{, }\StringTok{"email"}\NormalTok{: }\StringTok{"Email"}\NormalTok{\}}
\NormalTok{clean }\OperatorTok{=}\NormalTok{ complaints.assign(}
\NormalTok{    channel}\OperatorTok{=}\KeywordTok{lambda}\NormalTok{ d: (}
\NormalTok{        d[}\StringTok{"channel"}\NormalTok{]}
\NormalTok{        .astype(}\StringTok{"string"}\NormalTok{)}
\NormalTok{        .}\BuiltInTok{str}\NormalTok{.strip()}
\NormalTok{        .}\BuiltInTok{str}\NormalTok{.lower()}
\NormalTok{        .}\BuiltInTok{map}\NormalTok{(channel\_map)}
\NormalTok{        .fillna(}\StringTok{"Other"}\NormalTok{)}
\NormalTok{    ),}
\NormalTok{    opened\_month}\OperatorTok{=}\KeywordTok{lambda}\NormalTok{ d: d[}\StringTok{"opened\_at"}\NormalTok{].dt.to\_period(}\StringTok{"M"}\NormalTok{),}
\NormalTok{    resolution\_days}\OperatorTok{=}\KeywordTok{lambda}\NormalTok{ d: (d[}\StringTok{"closed\_at"}\NormalTok{] }\OperatorTok{{-}}\NormalTok{ d[}\StringTok{"opened\_at"}\NormalTok{]).dt.days,}
\NormalTok{    overdue}\OperatorTok{=}\KeywordTok{lambda}\NormalTok{ d: (}
\NormalTok{        d[}\StringTok{"closed\_at"}\NormalTok{].isna()}
        \OperatorTok{\&}\NormalTok{ (pd.Timestamp.today() }\OperatorTok{{-}}\NormalTok{ d[}\StringTok{"opened\_at"}\NormalTok{]).dt.days.gt(d[}\StringTok{"sla\_days"}\NormalTok{])}
\NormalTok{    ),}
\NormalTok{)}

\NormalTok{mgmt\_summary }\OperatorTok{=}\NormalTok{ clean.groupby([}\StringTok{"opened\_month"}\NormalTok{, }\StringTok{"product"}\NormalTok{, }\StringTok{"severity"}\NormalTok{], observed}\OperatorTok{=}\VariableTok{True}\NormalTok{).agg(}
\NormalTok{    cases}\OperatorTok{=}\NormalTok{(}\StringTok{"case\_id"}\NormalTok{, }\StringTok{"nunique"}\NormalTok{),}
\NormalTok{    median\_resolution\_days}\OperatorTok{=}\NormalTok{(}\StringTok{"resolution\_days"}\NormalTok{, }\StringTok{"median"}\NormalTok{),}
\NormalTok{    open\_cases}\OperatorTok{=}\NormalTok{(}\StringTok{"closed\_at"}\NormalTok{, }\KeywordTok{lambda}\NormalTok{ s: s.isna().}\BuiltInTok{sum}\NormalTok{()),}
\NormalTok{    overdue\_cases}\OperatorTok{=}\NormalTok{(}\StringTok{"overdue"}\NormalTok{, }\StringTok{"sum"}\NormalTok{),}
\NormalTok{).reset\_index()}

\NormalTok{overdue\_cases }\OperatorTok{=}\NormalTok{ clean.loc[clean[}\StringTok{"overdue"}\NormalTok{]].sort\_values(}\StringTok{"opened\_at"}\NormalTok{)}
\end{Highlighting}
\end{Shaded}

\subsection{6.5 Example 5: SQL-to-Pandas
Workflow}\label{example-5-sql-to-pandas-workflow}

In many organizations the database remains the authoritative store for
transactional data, while \texttt{pandas} is used for validation,
reshaping, and analytical features that are easier to express in Python.
This example pulls a bounded set of order rows from PostgreSQL with
\texttt{read\_sql()}, computes daily revenue and order counts in memory,
reconciles the daily aggregate against the detail extract, and writes
the cleaned summary back to the database with \texttt{to\_sql()}.

\begin{Shaded}
\begin{Highlighting}[]
\ImportTok{import}\NormalTok{ pandas }\ImportTok{as}\NormalTok{ pd}
\ImportTok{from}\NormalTok{ sqlalchemy }\ImportTok{import}\NormalTok{ create\_engine}

\NormalTok{engine }\OperatorTok{=}\NormalTok{ create\_engine(}\StringTok{"postgresql+psycopg2://user:pass@localhost/analytics"}\NormalTok{)}
\NormalTok{sales }\OperatorTok{=}\NormalTok{ pd.read\_sql(}
    \StringTok{"""}
\StringTok{    SELECT o.order\_id, o.customer\_id, o.order\_date, o.quantity, o.unit\_price}
\StringTok{    FROM sales.orders o}
\StringTok{    WHERE o.order\_date \textgreater{}= }\SpecialCharTok{\%(start)s}\StringTok{ AND o.order\_date \textless{} }\SpecialCharTok{\%(end)s}
\StringTok{    """}\NormalTok{,}
\NormalTok{    engine,}
\NormalTok{    params}\OperatorTok{=}\NormalTok{\{}\StringTok{"start"}\NormalTok{: }\StringTok{"2025{-}01{-}01"}\NormalTok{, }\StringTok{"end"}\NormalTok{: }\StringTok{"2025{-}02{-}01"}\NormalTok{\},}
\NormalTok{    dtype}\OperatorTok{=}\NormalTok{\{}\StringTok{"order\_id"}\NormalTok{: }\StringTok{"string"}\NormalTok{, }\StringTok{"customer\_id"}\NormalTok{: }\StringTok{"string"}\NormalTok{\},}
\NormalTok{    parse\_dates}\OperatorTok{=}\NormalTok{[}\StringTok{"order\_date"}\NormalTok{],}
\NormalTok{)}

\NormalTok{sales }\OperatorTok{=}\NormalTok{ sales.assign(revenue}\OperatorTok{=}\KeywordTok{lambda}\NormalTok{ d: d[}\StringTok{"quantity"}\NormalTok{] }\OperatorTok{*}\NormalTok{ d[}\StringTok{"unit\_price"}\NormalTok{])}
\NormalTok{daily }\OperatorTok{=}\NormalTok{ (}
\NormalTok{    sales}
\NormalTok{    .groupby(sales[}\StringTok{"order\_date"}\NormalTok{].dt.date)}
\NormalTok{    .agg(revenue}\OperatorTok{=}\NormalTok{(}\StringTok{"revenue"}\NormalTok{, }\StringTok{"sum"}\NormalTok{), orders}\OperatorTok{=}\NormalTok{(}\StringTok{"order\_id"}\NormalTok{, }\StringTok{"nunique"}\NormalTok{))}
\NormalTok{    .reset\_index()}
\NormalTok{)}

\ControlFlowTok{assert}\NormalTok{ daily[}\StringTok{"revenue"}\NormalTok{].}\BuiltInTok{sum}\NormalTok{() }\OperatorTok{==}\NormalTok{ sales[}\StringTok{"revenue"}\NormalTok{].}\BuiltInTok{sum}\NormalTok{()}
\NormalTok{daily.to\_sql(}\StringTok{"daily\_revenue\_clean"}\NormalTok{, engine, if\_exists}\OperatorTok{=}\StringTok{"replace"}\NormalTok{, index}\OperatorTok{=}\VariableTok{False}\NormalTok{)}
\end{Highlighting}
\end{Shaded}

\subsection{6.6 Example 6: Time Series
Monitoring}\label{example-6-time-series-monitoring}

Operations and quality teams often monitor sensors, transaction volumes,
or other metrics that arrive as time-stamped readings. Raw event-level
data are too granular for daily review, so analysts need rollups,
smoothed baselines, and simple anomaly rules. This workflow resamples
readings to daily averages by production line, computes a seven-day
rolling mean, flags days that exceed two standard deviations above the
local baseline, and summarizes how many anomalies occurred per line.

\begin{Shaded}
\begin{Highlighting}[]
\ImportTok{import}\NormalTok{ numpy }\ImportTok{as}\NormalTok{ np}
\ImportTok{import}\NormalTok{ pandas }\ImportTok{as}\NormalTok{ pd}

\NormalTok{ops }\OperatorTok{=}\NormalTok{ pd.read\_csv(}\StringTok{"sensor\_readings.csv"}\NormalTok{, parse\_dates}\OperatorTok{=}\NormalTok{[}\StringTok{"timestamp"}\NormalTok{])}
\NormalTok{ops }\OperatorTok{=}\NormalTok{ ops.set\_index(}\StringTok{"timestamp"}\NormalTok{).sort\_index()}

\NormalTok{daily }\OperatorTok{=}\NormalTok{ ops.groupby(}\StringTok{"line"}\NormalTok{).resample(}\StringTok{"D"}\NormalTok{).agg(}
\NormalTok{    avg\_value}\OperatorTok{=}\NormalTok{(}\StringTok{"value"}\NormalTok{, }\StringTok{"mean"}\NormalTok{),}
\NormalTok{    max\_value}\OperatorTok{=}\NormalTok{(}\StringTok{"value"}\NormalTok{, }\StringTok{"max"}\NormalTok{),}
\NormalTok{    readings}\OperatorTok{=}\NormalTok{(}\StringTok{"value"}\NormalTok{, }\StringTok{"count"}\NormalTok{),}
\NormalTok{).reset\_index()}

\NormalTok{daily[}\StringTok{"seven\_day\_avg"}\NormalTok{] }\OperatorTok{=}\NormalTok{ (}
\NormalTok{    daily.sort\_values([}\StringTok{"line"}\NormalTok{, }\StringTok{"timestamp"}\NormalTok{])}
\NormalTok{    .groupby(}\StringTok{"line"}\NormalTok{, observed}\OperatorTok{=}\VariableTok{True}\NormalTok{)[}\StringTok{"avg\_value"}\NormalTok{]}
\NormalTok{    .transform(}\KeywordTok{lambda}\NormalTok{ s: s.rolling(}\DecValTok{7}\NormalTok{, min\_periods}\OperatorTok{=}\DecValTok{3}\NormalTok{).mean())}
\NormalTok{)}
\NormalTok{daily[}\StringTok{"anomaly"}\NormalTok{] }\OperatorTok{=}\NormalTok{ daily[}\StringTok{"avg\_value"}\NormalTok{] }\OperatorTok{\textgreater{}}\NormalTok{ (}
\NormalTok{    daily[}\StringTok{"seven\_day\_avg"}\NormalTok{]}
    \OperatorTok{+} \DecValTok{2} \OperatorTok{*}\NormalTok{ daily.groupby(}\StringTok{"line"}\NormalTok{)[}\StringTok{"avg\_value"}\NormalTok{].transform(}\StringTok{"std"}\NormalTok{)}
\NormalTok{)}

\NormalTok{monitoring\_report }\OperatorTok{=}\NormalTok{ (}
\NormalTok{    daily}
\NormalTok{    .groupby([}\StringTok{"line"}\NormalTok{, }\StringTok{"anomaly"}\NormalTok{], observed}\OperatorTok{=}\VariableTok{True}\NormalTok{)}
\NormalTok{    .size()}
\NormalTok{    .unstack(fill\_value}\OperatorTok{=}\DecValTok{0}\NormalTok{)}
\NormalTok{)}
\end{Highlighting}
\end{Shaded}

\newpage

\subsection{6.7 Example 7: Machine Learning
Preparation}\label{example-7-machine-learning-preparation}

Before fitting a classifier, analysts must define features explicitly,
separate numeric and categorical columns, and keep preprocessing
consistent between training and evaluation. This example starts from the
enriched sales table, creates a binary target for high-margin orders,
lists feature columns by name, and builds a scikit-learn
\texttt{Pipeline} with imputation, scaling, one-hot encoding, and a
random forest evaluated by cross-validation with a fixed random seed.

\begin{Shaded}
\begin{Highlighting}[]
\ImportTok{from}\NormalTok{ sklearn.compose }\ImportTok{import}\NormalTok{ ColumnTransformer}
\ImportTok{from}\NormalTok{ sklearn.ensemble }\ImportTok{import}\NormalTok{ RandomForestClassifier}
\ImportTok{from}\NormalTok{ sklearn.impute }\ImportTok{import}\NormalTok{ SimpleImputer}
\ImportTok{from}\NormalTok{ sklearn.model\_selection }\ImportTok{import}\NormalTok{ cross\_val\_score}
\ImportTok{from}\NormalTok{ sklearn.pipeline }\ImportTok{import}\NormalTok{ Pipeline}
\ImportTok{from}\NormalTok{ sklearn.preprocessing }\ImportTok{import}\NormalTok{ OneHotEncoder, StandardScaler}

\NormalTok{model\_data }\OperatorTok{=}\NormalTok{ enriched.assign(}
\NormalTok{    high\_margin\_order}\OperatorTok{=}\KeywordTok{lambda}\NormalTok{ d: ((d[}\StringTok{"gross\_margin"}\NormalTok{] }\OperatorTok{/}\NormalTok{ d[}\StringTok{"revenue"}\NormalTok{]) }\OperatorTok{\textgreater{}=} \FloatTok{0.40}\NormalTok{).astype(}\BuiltInTok{int}\NormalTok{),}
\NormalTok{    order\_month}\OperatorTok{=}\KeywordTok{lambda}\NormalTok{ d: d[}\StringTok{"order\_date"}\NormalTok{].dt.month,}
\NormalTok{)}

\NormalTok{feature\_columns }\OperatorTok{=}\NormalTok{ [}\StringTok{"quantity"}\NormalTok{, }\StringTok{"unit\_price"}\NormalTok{, }\StringTok{"order\_month"}\NormalTok{, }\StringTok{"region"}\NormalTok{, }\StringTok{"segment"}\NormalTok{, }\StringTok{"category"}\NormalTok{]}
\NormalTok{target\_column }\OperatorTok{=} \StringTok{"high\_margin\_order"}
\NormalTok{X }\OperatorTok{=}\NormalTok{ model\_data[feature\_columns]}
\NormalTok{y }\OperatorTok{=}\NormalTok{ model\_data[target\_column]}

\NormalTok{numeric\_features }\OperatorTok{=}\NormalTok{ [}\StringTok{"quantity"}\NormalTok{, }\StringTok{"unit\_price"}\NormalTok{, }\StringTok{"order\_month"}\NormalTok{]}
\NormalTok{categorical\_features }\OperatorTok{=}\NormalTok{ [}\StringTok{"region"}\NormalTok{, }\StringTok{"segment"}\NormalTok{, }\StringTok{"category"}\NormalTok{]}

\NormalTok{preprocess }\OperatorTok{=}\NormalTok{ ColumnTransformer([}
\NormalTok{    (}
        \StringTok{"numeric"}\NormalTok{,}
\NormalTok{        Pipeline([}
\NormalTok{            (}\StringTok{"impute"}\NormalTok{, SimpleImputer(strategy}\OperatorTok{=}\StringTok{"median"}\NormalTok{)),}
\NormalTok{            (}\StringTok{"scale"}\NormalTok{, StandardScaler()),}
\NormalTok{        ]),}
\NormalTok{        numeric\_features,}
\NormalTok{    ),}
\NormalTok{    (}
        \StringTok{"categorical"}\NormalTok{,}
\NormalTok{        Pipeline([}
\NormalTok{            (}\StringTok{"impute"}\NormalTok{, SimpleImputer(strategy}\OperatorTok{=}\StringTok{"most\_frequent"}\NormalTok{)),}
\NormalTok{            (}\StringTok{"encode"}\NormalTok{, OneHotEncoder(handle\_unknown}\OperatorTok{=}\StringTok{"ignore"}\NormalTok{)),}
\NormalTok{        ]),}
\NormalTok{        categorical\_features,}
\NormalTok{    ),}
\NormalTok{])}

\NormalTok{classifier }\OperatorTok{=}\NormalTok{ Pipeline([}
\NormalTok{    (}\StringTok{"preprocess"}\NormalTok{, preprocess),}
\NormalTok{    (}
        \StringTok{"model"}\NormalTok{,}
\NormalTok{        RandomForestClassifier(n\_estimators}\OperatorTok{=}\DecValTok{200}\NormalTok{, random\_state}\OperatorTok{=}\DecValTok{42}\NormalTok{),}
\NormalTok{    ),}
\NormalTok{])}

\NormalTok{scores }\OperatorTok{=}\NormalTok{ cross\_val\_score(classifier, X, y, cv}\OperatorTok{=}\DecValTok{3}\NormalTok{, scoring}\OperatorTok{=}\StringTok{"accuracy"}\NormalTok{)}
\end{Highlighting}
\end{Shaded}

Keeping \texttt{feature\_columns} explicit preserves interpretability
and reduces leakage from ad hoc column drops.

\section{7. From Notebook Exploration to Reproducible
Workflow}\label{from-notebook-exploration-to-reproducible-workflow}

Notebooks are excellent for exploration because they keep code,
narrative, intermediate tables, figures, and interpretation in one
visible workspace. This matters in \texttt{pandas} work because early
data analysis is often uncertain: the analyst may need to inspect
unexpected values, test several cleaning rules, compare alternative
joins, and show provisional outputs to collaborators before the final
transformation logic is known. Recent studies of computational notebooks
emphasize this exploratory value. Scientists and data analysts use
notebooks to combine explanation and execution, to preserve the
reasoning around intermediate results, and to move fluidly between data
inspection and interpretation {[}13, 21, 23{]}. In that stage, the
notebook is not a weakness. It is a productive interface for discovery.

The same features become risky when a notebook is treated as the sole
system of record for a recurring analytical job. Notebook execution is
stateful: cells can be run out of order, variables can persist after the
code that created them has been edited or deleted, and displayed output
may not correspond to a clean top-to-bottom rerun. Empirical work on
Jupyter notebooks has repeatedly identified reproducibility problems
associated with hidden state, ambiguous execution order, missing
dependencies, hard-coded paths, and environment drift {[}19, 20{]}.
Studies of publication-associated notebooks show that these risks are
not confined to casual analysis; they also appear in scientific
workflows where computational reproducibility is part of the scholarly
record {[}22{]}. For business analytics, the practical consequence is
similar: a notebook that produced last month's dashboard may not
reliably explain, reproduce, or validate this month's report.

The appropriate response is not to ban notebooks or pretend that all
analysis should begin as production software. The better response is to
assign notebooks a clear role in the workflow lifecycle. Notebooks are
well suited to asking questions of the data, documenting exploratory
reasoning, and exposing anomalies that were not visible in the original
spreadsheet or extract. Once a transformation becomes stable, repeated,
or decision-relevant, the logic should be promoted into named functions,
scripts, tests, and validation checks. This transition is especially
important for \texttt{pandas} because the most consequential steps are
often small table operations: a dtype declaration that preserves leading
zeros, a join validation that prevents duplicate-key inflation, a filter
that removes subtotal rows, or an assertion that blocks report export
when totals do not reconcile.

A practical maturity path therefore treats notebooks as exploratory
workspaces and treats scripts as the durable record of repeated
analysis. The notebook may remain in the project as a narrative
artifact, but the pipeline should be runnable from a clean process with
declared inputs, pinned dependencies, immutable raw data, logged row
counts, and explicit failure conditions. This division of labor
preserves the strengths of notebooks while reducing the audit and
reproducibility risks identified in recent notebook studies {[}19, 20,
22, 23{]}. Figure 7 and Table 5 summarize this maturity path.

\textbf{Figure 7.} Maturity path from exploratory notebooks to
reproducible project workflows.

\begin{center}
\includegraphics[width=0.96\linewidth,height=0.62\textheight,keepaspectratio]{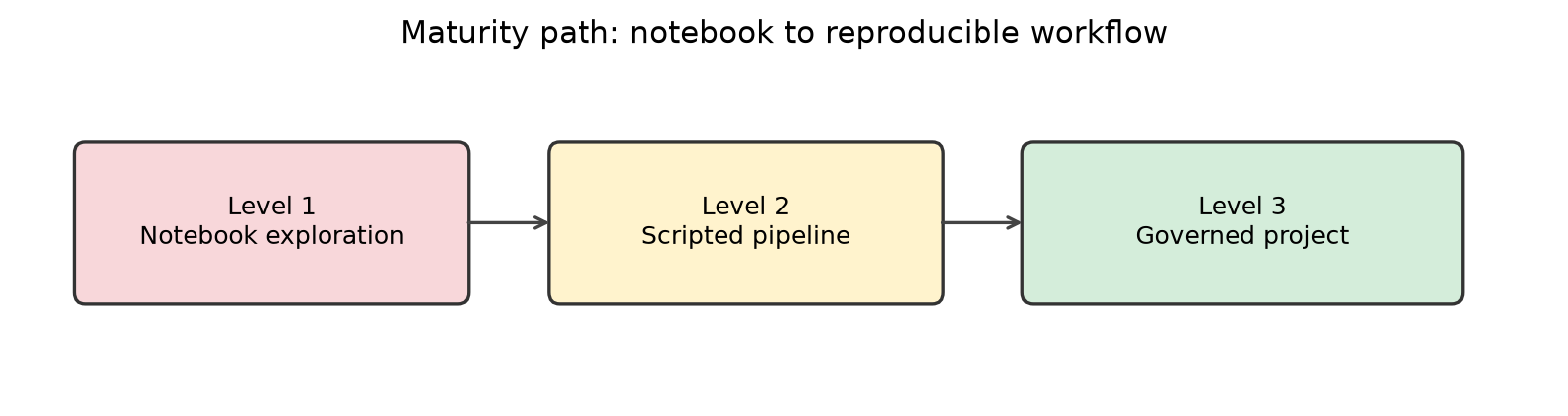}
\end{center}

\textbf{Table 5.} Reproducibility principles and implementation
practices

{\def\LTcaptype{none} 
\begin{longtable}[]{@{}
  >{\raggedright\arraybackslash}p{(\linewidth - 4\tabcolsep) * \real{0.3333}}
  >{\raggedright\arraybackslash}p{(\linewidth - 4\tabcolsep) * \real{0.3333}}
  >{\raggedright\arraybackslash}p{(\linewidth - 4\tabcolsep) * \real{0.3333}}@{}}
\toprule\noalign{}
\begin{minipage}[b]{\linewidth}\raggedright
Principle
\end{minipage} & \begin{minipage}[b]{\linewidth}\raggedright
Implementation practice
\end{minipage} & \begin{minipage}[b]{\linewidth}\raggedright
Example artifact
\end{minipage} \\
\midrule\noalign{}
\endhead
\bottomrule\noalign{}
\endlastfoot
Raw data immutable & Copy once; never edit in place &
\texttt{data/raw/} \\
Scripted processing & Write only from successful runs &
\texttt{data/processed/} \\
Notebooks for exploration & Promote stable cells to modules & notebook
to script \\
Dependencies pinned & Record package versions &
\texttt{requirements.txt} \\
Randomness controlled & Set seeds in models &
\texttt{random\_state=42} \\
Outputs traceable & Log timestamps and row counts & run ID and row
count \\
Assumptions tested & Fail pipeline on violation & uniqueness
assertion \\
\end{longtable}
}

\textbf{Principles:}

\begin{itemize}
\tightlist
\item
  \textbf{Raw data should be immutable.} Copy extracts into
  \texttt{data/raw/} and never edit them in place.
\item
  \textbf{Processed outputs should be written by scripts.} Save parquet
  or CSV snapshots to \texttt{data/processed/} only after
  transformations succeed.
\item
  \textbf{Scripts beat notebooks for repetition.} Monthly close, survey
  refreshes, and compliance reports should run as named entry points.
\item
  \textbf{Pin dependency versions.} Use \texttt{requirements.txt} or
  \texttt{environment.yml} and record the environment used for published
  results {[}10{]}.
\item
  \textbf{Fix random seeds for modeling.} Set \texttt{random\_state} in
  sklearn and NumPy when results must be reproduced.
\item
  \textbf{Attach metadata to outputs.} Include run timestamp, git commit
  hash, and row counts in logs or report tabs.
\item
  \textbf{Validate assumptions with assertions.} Treat failed checks as
  pipeline failures, not warnings buried in output.
\end{itemize}

\textbf{Example project structure:}

\begin{Shaded}
\begin{Highlighting}[]
\NormalTok{project/}
\NormalTok{  data/}
\NormalTok{    raw/}
\NormalTok{    interim/}
\NormalTok{    processed/}
\NormalTok{  notebooks/}
\NormalTok{  src/}
\NormalTok{    cleaning.py}
\NormalTok{    validation.py}
\NormalTok{    reporting.py}
\NormalTok{  reports/}
\NormalTok{  tests/}
\NormalTok{  requirements.txt}
\NormalTok{  README.md}
\end{Highlighting}
\end{Shaded}

\textbf{Logging intermediate outputs and row counts:}

\begin{Shaded}
\begin{Highlighting}[]
\ImportTok{import}\NormalTok{ logging}
\ImportTok{from}\NormalTok{ datetime }\ImportTok{import}\NormalTok{ datetime, timezone}
\ImportTok{from}\NormalTok{ pathlib }\ImportTok{import}\NormalTok{ Path}
\ImportTok{import}\NormalTok{ pandas }\ImportTok{as}\NormalTok{ pd}

\NormalTok{logging.basicConfig(level}\OperatorTok{=}\NormalTok{logging.INFO, }\BuiltInTok{format}\OperatorTok{=}\StringTok{"}\SpecialCharTok{\%(asctime)s}\StringTok{ }\SpecialCharTok{\%(levelname)s}\StringTok{ }\SpecialCharTok{\%(message)s}\StringTok{"}\NormalTok{)}
\NormalTok{RUN\_ID }\OperatorTok{=}\NormalTok{ datetime.now(timezone.utc).strftime(}\StringTok{"\%Y\%m}\SpecialCharTok{\%d}\StringTok{T\%H\%M\%SZ"}\NormalTok{)}

\KeywordTok{def}\NormalTok{ save\_stage(df: pd.DataFrame, name: }\BuiltInTok{str}\NormalTok{, interim\_dir: Path) }\OperatorTok{{-}\textgreater{}}\NormalTok{ pd.DataFrame:}
\NormalTok{    path }\OperatorTok{=}\NormalTok{ interim\_dir }\OperatorTok{/} \SpecialStringTok{f"}\SpecialCharTok{\{}\NormalTok{RUN\_ID}\SpecialCharTok{\}}\SpecialStringTok{\_}\SpecialCharTok{\{}\NormalTok{name}\SpecialCharTok{\}}\SpecialStringTok{.parquet"}
\NormalTok{    df.to\_parquet(path, index}\OperatorTok{=}\VariableTok{False}\NormalTok{)}
\NormalTok{    logging.info(}\StringTok{"}\SpecialCharTok{\%s}\StringTok{: }\SpecialCharTok{\%s}\StringTok{ rows {-}\textgreater{} }\SpecialCharTok{\%s}\StringTok{"}\NormalTok{, name, }\SpecialStringTok{f"}\SpecialCharTok{\{}\BuiltInTok{len}\NormalTok{(df)}\SpecialCharTok{:,\}}\SpecialStringTok{"}\NormalTok{, path)}
    \ControlFlowTok{return}\NormalTok{ df}

\NormalTok{interim }\OperatorTok{=}\NormalTok{ Path(}\StringTok{"data/interim"}\NormalTok{)}
\NormalTok{orders }\OperatorTok{=}\NormalTok{ save\_stage(}
\NormalTok{    pd.read\_excel(}\StringTok{"data/raw/orders.xlsx"}\NormalTok{, dtype}\OperatorTok{=}\NormalTok{\{}\StringTok{"order\_id"}\NormalTok{: }\StringTok{"string"}\NormalTok{\}),}
    \StringTok{"orders\_raw"}\NormalTok{,}
\NormalTok{    interim,}
\NormalTok{)}
\NormalTok{clean\_orders }\OperatorTok{=}\NormalTok{ save\_stage(orders.dropna(subset}\OperatorTok{=}\NormalTok{[}\StringTok{"order\_id"}\NormalTok{]), }\StringTok{"orders\_clean"}\NormalTok{, interim)}
\end{Highlighting}
\end{Shaded}

\section{8. Common Failure Modes in Excel-to-Pandas
Migration}\label{common-failure-modes-in-excel-to-pandas-migration}

Excel habits do not always transfer safely. The following failure modes
appear frequently in teaching and consulting practice. Table 6 provides
a quick reference; Sections 8.1--8.12 expand each mode with recommended
pandas patterns.

\textbf{Table 6.} Common Excel-to-pandas failure modes

{\def\LTcaptype{none} 
\begin{longtable}[]{@{}
  >{\raggedright\arraybackslash}p{(\linewidth - 6\tabcolsep) * \real{0.2500}}
  >{\raggedright\arraybackslash}p{(\linewidth - 6\tabcolsep) * \real{0.2500}}
  >{\raggedright\arraybackslash}p{(\linewidth - 6\tabcolsep) * \real{0.2500}}
  >{\raggedright\arraybackslash}p{(\linewidth - 6\tabcolsep) * \real{0.2500}}@{}}
\toprule\noalign{}
\begin{minipage}[b]{\linewidth}\raggedright
\#
\end{minipage} & \begin{minipage}[b]{\linewidth}\raggedright
Failure mode
\end{minipage} & \begin{minipage}[b]{\linewidth}\raggedright
Symptom
\end{minipage} & \begin{minipage}[b]{\linewidth}\raggedright
Recommended pattern
\end{minipage} \\
\midrule\noalign{}
\endhead
\bottomrule\noalign{}
\endlastfoot
1 & Losing leading zeros & IDs change (\texttt{00421} to \texttt{421}) &
Read IDs as strings \\
2 & Formatted dates as strings & Wrong sort order & Parse dates
explicitly \\
3 & Dirty join keys & Unexpected nulls after merge & Strip and normalize
case \\
4 & Duplicate lookup keys & Inflated totals & Validate merge
cardinality \\
5 & Empty string vs missing & Bad aggregates & Convert blanks to
missing \\
6 & Hidden subtotal rows & Double-counted metrics & Filter
\texttt{total} / \texttt{subtotal} rows \\
7 & Mixed numeric/text & \texttt{object} dtype columns & Parse currency
to \texttt{Float64} \\
8 & Out-of-order notebook cells & State differs from displayed code &
Scripts or executed notebooks \\
9 & Chained assignment & Silent failures / CoW errors & Assign with
\texttt{.loc} \\
10 & Unpinned package versions & Drift across machines & Pinned
\texttt{requirements.txt} \\
11 & Export before validation & Bad numbers in reports & Assert before
export \\
12 & Treating pandas as a database & Memory / concurrency issues & SQL
or warehouse for authoritative store \\
\end{longtable}
}

\subsection{8.1 Losing Leading Zeros in
IDs}\label{losing-leading-zeros-in-ids}

Excel often stores IDs as numbers; CSV export turns \texttt{00421} into
\texttt{421}.

\begin{Shaded}
\begin{Highlighting}[]
\NormalTok{df }\OperatorTok{=}\NormalTok{ pd.read\_csv(}\StringTok{"customers.csv"}\NormalTok{)}

\NormalTok{df }\OperatorTok{=}\NormalTok{ pd.read\_csv(}\StringTok{"customers.csv"}\NormalTok{, dtype}\OperatorTok{=}\NormalTok{\{}\StringTok{"customer\_id"}\NormalTok{: }\StringTok{"string"}\NormalTok{\})}
\end{Highlighting}
\end{Shaded}

\subsection{8.2 Treating Formatted Dates as Valid
Dates}\label{treating-formatted-dates-as-valid-dates}

A column can look like dates in Excel while remaining strings in
\texttt{pandas}.

\begin{Shaded}
\begin{Highlighting}[]
\NormalTok{df.sort\_values(}\StringTok{"order\_date"}\NormalTok{)}

\NormalTok{df }\OperatorTok{=}\NormalTok{ df.assign(order\_date}\OperatorTok{=}\NormalTok{pd.to\_datetime(df[}\StringTok{"order\_date"}\NormalTok{], errors}\OperatorTok{=}\StringTok{"coerce"}\NormalTok{))}
\ControlFlowTok{assert}\NormalTok{ df[}\StringTok{"order\_date"}\NormalTok{].notna().}\BuiltInTok{all}\NormalTok{()}
\end{Highlighting}
\end{Shaded}

\subsection{8.3 Joining on Dirty Keys}\label{joining-on-dirty-keys}

Hidden spaces and inconsistent casing break VLOOKUP-style joins.

\begin{Shaded}
\begin{Highlighting}[]
\NormalTok{keys }\OperatorTok{=}\NormalTok{ df.assign(}
\NormalTok{    customer\_id}\OperatorTok{=}\KeywordTok{lambda}\NormalTok{ d: d[}\StringTok{"customer\_id"}\NormalTok{].astype(}\StringTok{"string"}\NormalTok{).}\BuiltInTok{str}\NormalTok{.strip().}\BuiltInTok{str}\NormalTok{.upper()}
\NormalTok{)}
\NormalTok{lookup }\OperatorTok{=}\NormalTok{ lookup.assign(}
\NormalTok{    customer\_id}\OperatorTok{=}\KeywordTok{lambda}\NormalTok{ d: d[}\StringTok{"customer\_id"}\NormalTok{].astype(}\StringTok{"string"}\NormalTok{).}\BuiltInTok{str}\NormalTok{.strip().}\BuiltInTok{str}\NormalTok{.upper()}
\NormalTok{)}
\NormalTok{merged }\OperatorTok{=}\NormalTok{ keys.merge(lookup, on}\OperatorTok{=}\StringTok{"customer\_id"}\NormalTok{, how}\OperatorTok{=}\StringTok{"left"}\NormalTok{, validate}\OperatorTok{=}\StringTok{"m:1"}\NormalTok{)}
\end{Highlighting}
\end{Shaded}

\subsection{8.4 Accidentally Multiplying Rows in
Joins}\label{accidentally-multiplying-rows-in-joins}

Duplicate keys in a lookup table inflate totals silently in Excel;
\texttt{validate=} catches this in \texttt{pandas}.

\begin{Shaded}
\begin{Highlighting}[]
\NormalTok{enriched }\OperatorTok{=}\NormalTok{ orders.merge(customers, on}\OperatorTok{=}\StringTok{"customer\_id"}\NormalTok{, how}\OperatorTok{=}\StringTok{"left"}\NormalTok{, validate}\OperatorTok{=}\StringTok{"m:1"}\NormalTok{)}
\end{Highlighting}
\end{Shaded}

\subsection{8.5 Confusing Missing Values with Empty
Strings}\label{confusing-missing-values-with-empty-strings}

Excel blanks may become \texttt{""} rather than \texttt{pd.NA}, breaking
numeric aggregation.

\begin{Shaded}
\begin{Highlighting}[]
\NormalTok{df }\OperatorTok{=}\NormalTok{ df.replace(}\VerbatimStringTok{r"}\DecValTok{\^{}\textbackslash{}s}\OperatorTok{*}\DecValTok{$}\VerbatimStringTok{"}\NormalTok{, pd.NA, regex}\OperatorTok{=}\VariableTok{True}\NormalTok{)}
\ControlFlowTok{assert}\NormalTok{ df[}\StringTok{"revenue"}\NormalTok{].notna().}\BuiltInTok{all}\NormalTok{()}
\end{Highlighting}
\end{Shaded}

\subsection{8.6 Hidden Subtotal Rows}\label{hidden-subtotal-rows}

Subtotal rows embedded in detail sheets double-count in pivots.

\begin{Shaded}
\begin{Highlighting}[]
\NormalTok{detail }\OperatorTok{=}\NormalTok{ raw.loc[}\OperatorTok{\textasciitilde{}}\NormalTok{raw[}\StringTok{"order\_id"}\NormalTok{].astype(}\StringTok{"string"}\NormalTok{).}\BuiltInTok{str}\NormalTok{.lower().isin([}\StringTok{"total"}\NormalTok{, }\StringTok{"subtotal"}\NormalTok{])]}
\end{Highlighting}
\end{Shaded}

\subsection{8.7 Mixed Numeric and Text
Columns}\label{mixed-numeric-and-text-columns}

Currency symbols and thousands separators produce \texttt{object} dtype
columns.

\begin{Shaded}
\begin{Highlighting}[]
\NormalTok{df }\OperatorTok{=}\NormalTok{ df.assign(}
\NormalTok{    amount}\OperatorTok{=}\KeywordTok{lambda}\NormalTok{ d: (}
\NormalTok{        d[}\StringTok{"amount"}\NormalTok{]}
\NormalTok{        .astype(}\StringTok{"string"}\NormalTok{)}
\NormalTok{        .}\BuiltInTok{str}\NormalTok{.replace(}\VerbatimStringTok{r"}\PreprocessorTok{[$,]}\VerbatimStringTok{"}\NormalTok{, }\StringTok{""}\NormalTok{, regex}\OperatorTok{=}\VariableTok{True}\NormalTok{)}
\NormalTok{        .astype(}\StringTok{"Float64"}\NormalTok{)}
\NormalTok{    )}
\NormalTok{)}
\end{Highlighting}
\end{Shaded}

\subsection{8.8 Copying Notebook Cells Out of
Order}\label{copying-notebook-cells-out-of-order}

Notebook state can disagree with displayed code. Prefer
\texttt{nbconvert\ -\/-execute} or scripts for recurring jobs.

\subsection{8.9 Using Chained Assignment
Incorrectly}\label{using-chained-assignment-incorrectly}

Chained indexing can fail silently or raise in Copy-on-Write mode
{[}1{]}.

\begin{Shaded}
\begin{Highlighting}[]
\NormalTok{df[df[}\StringTok{"region"}\NormalTok{].eq(}\StringTok{"West"}\NormalTok{)][}\StringTok{"revenue"}\NormalTok{] }\OperatorTok{=} \DecValTok{0}

\NormalTok{df.loc[df[}\StringTok{"region"}\NormalTok{].eq(}\StringTok{"West"}\NormalTok{), }\StringTok{"revenue"}\NormalTok{] }\OperatorTok{=} \DecValTok{0}
\end{Highlighting}
\end{Shaded}

\subsection{8.10 Failing to Pin Package
Versions}\label{failing-to-pin-package-versions}

Minor \texttt{pandas} releases can change parsing or copy behavior. Pin
versions for long-lived pipelines.

\subsection{8.11 Exporting Reports Without Validation
Checks}\label{exporting-reports-without-validation-checks}

Writing Excel output before reconciliation repeats spreadsheet risk in a
new format.

\begin{Shaded}
\begin{Highlighting}[]
\ControlFlowTok{assert}\NormalTok{ np.isclose(detail[}\StringTok{"revenue"}\NormalTok{].}\BuiltInTok{sum}\NormalTok{(), summary[}\StringTok{"revenue"}\NormalTok{].}\BuiltInTok{sum}\NormalTok{())}
\NormalTok{monthly\_metrics.to\_excel(}\StringTok{"report.xlsx"}\NormalTok{, index}\OperatorTok{=}\VariableTok{False}\NormalTok{)}
\end{Highlighting}
\end{Shaded}

\subsection{8.12 Assuming Pandas Is a
Database}\label{assuming-pandas-is-a-database}

\texttt{pandas} holds tables in memory for analysis. It does not replace
transactional storage, concurrent writes, or fine-grained permissions.

\section{9. Integration with the Python Data
Ecosystem}\label{integration-with-the-python-data-ecosystem}

\texttt{pandas} is rarely used alone. Its value comes partly from the
way it connects tools.

NumPy provides efficient numerical arrays and mathematical operations.
\texttt{pandas} builds on this style of vectorized computation while
adding labels, heterogeneous columns, missing data handling, and tabular
operations {[}3{]}. Matplotlib and seaborn turn \texttt{DataFrame}
columns into exploratory and publication-ready visualizations {[}5{]}.
SciPy and statsmodels support statistical testing and modeling.
scikit-learn provides machine learning pipelines that can consume
\texttt{pandas} data while preserving feature names through much of the
workflow {[}4, 15{]}. SQL databases remain essential for storage,
permissions, and multi-user access; \texttt{pandas} can read query
results into \texttt{DataFrame} objects for analysis and write
transformed data back to databases {[}14{]}.

The practical architecture in many organizations is therefore hybrid
(Figure 8):

\begin{itemize}
\tightlist
\item
  source systems and databases hold authoritative records;
\item
  Excel remains a familiar input and output layer for some teams;
\item
  \texttt{pandas} scripts perform repeatable cleaning, joining,
  aggregation, and validation;
\item
  visualization and modeling libraries consume the prepared tables;
\item
  reports, dashboards, and workbook outputs communicate results.
\end{itemize}

\textbf{Figure 8.} Hybrid data architecture with pandas connecting
source systems, Excel interfaces, and downstream analytics.

\begin{center}
\includegraphics[width=0.96\linewidth,height=0.62\textheight,keepaspectratio]{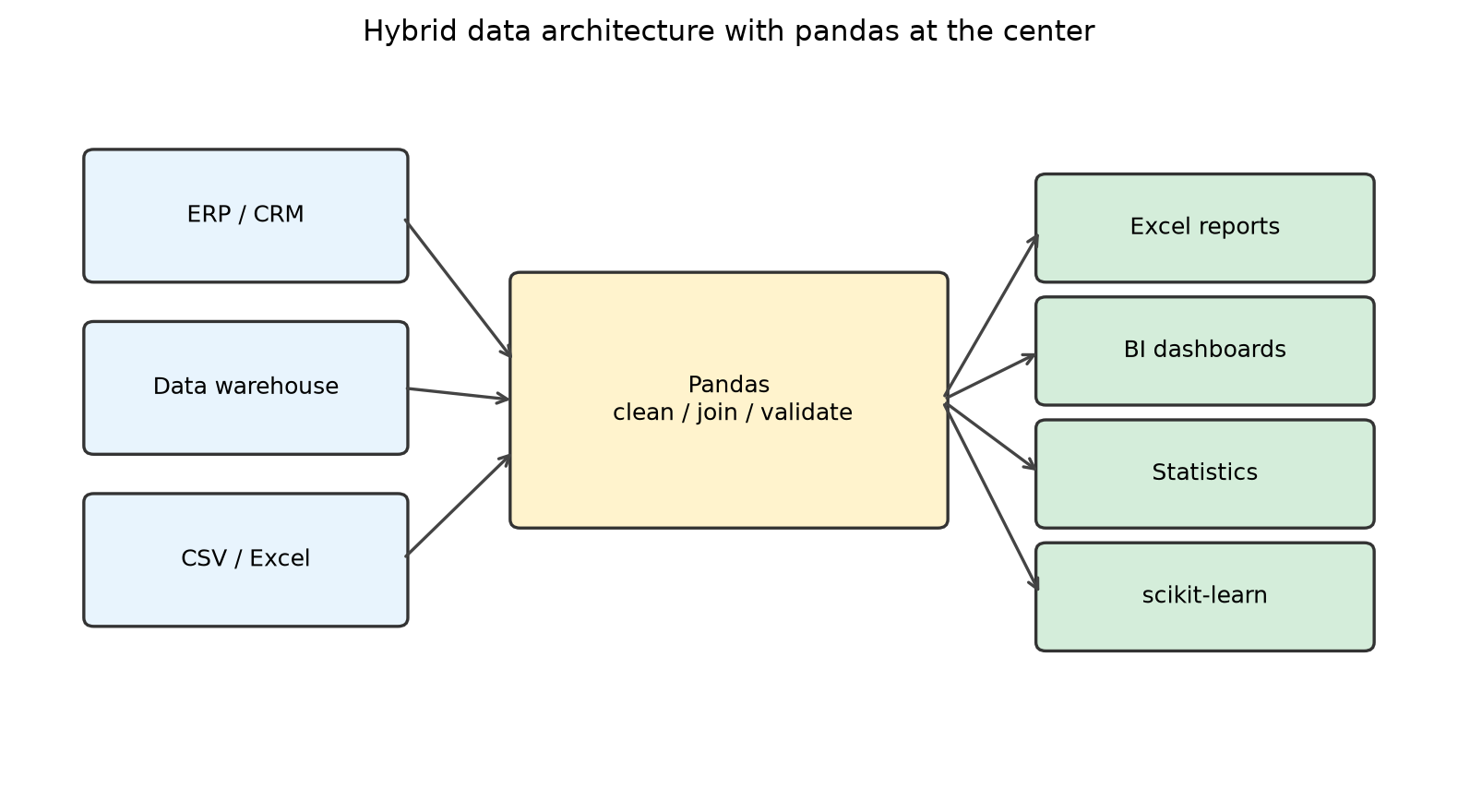}
\end{center}

This hybrid model is often more realistic than insisting on an immediate
end to spreadsheets. It allows organizations to reduce spreadsheet risk
while respecting existing stakeholder practices.

\section{10. Discussion}\label{discussion}

The workflows, examples, and failure modes presented in this paper
support a single practical conclusion: \texttt{pandas} is most valuable
when it functions as a disciplined middle layer between raw data and
analytical decisions. It does not make spreadsheets obsolete. Instead,
it gives analysts a way to encode cleaning, joining, validation, and
reporting in forms that can be reviewed, rerun, and handed to
collaborators without losing the table-oriented habits they already
possess. That conclusion is easy to state and harder to
institutionalize. The subsections below examine why the transition
matters from five complementary perspectives---organizational adoption,
research practice, business governance, teaching, and technical scale.

Across these perspectives, two themes recur. First, the workflow
taxonomy in Table 3 is not merely a catalog of functions. It is a way to
decompose messy real-world tasks into stages that can be owned, tested,
and improved independently. Second, the hybrid architecture in Figure 8
is a realistic picture of how many units already work: data enter from
databases and spreadsheets, transformation happens in code, and results
return to Excel, dashboards, or modeling tools. The discussion therefore
returns repeatedly to division of labor rather than tool replacement.

\subsection{10.1 Pandas as a Bridge Between Excel and Software
Engineering}\label{pandas-as-a-bridge-between-excel-and-software-engineering}

\texttt{pandas} lowers the cost of adopting table-oriented code because
it preserves familiar metaphors. Rows, columns, filters, pivots, and
lookup-style joins all have close analogues in the library. An analyst
who has spent years building pivot tables and VLOOKUP formulas can
usually understand what a \texttt{groupby()} or validated
\texttt{merge()} is trying to accomplish before they master Python
syntax in full. That familiarity matters in practice because adoption
failures in analytics teams often occur not when code is impossible to
learn, but when the first programming tools presented feel unrelated to
the work people already perform every month.

The bridge works in the other direction as well. Spreadsheets teach
important habits that remain relevant in code: attention to units,
skepticism about totals that look too clean, and awareness that a single
bad key can distort an entire table. \texttt{pandas} does not erase
those instincts. It gives them a more durable expression. When a student
asks why a join doubled revenue, the answer is visible in duplicate keys
rather than buried in a pasted range. When a manager asks how a regional
total was produced, the answer can be traced through named steps instead
of reconstructed from color-coded tabs.

At the same time, \texttt{pandas} introduces software engineering habits
that spreadsheets do not enforce on their own. Named functions, modular
files, version control, dependency pinning, and automated checks are not
optional extras for teams that expect the same report every month. The
examples in Sections 5 and 6 show how a monthly sales workbook can be
expressed as a short pipeline with explicit stages for ingestion,
enrichment, validation, and export. Once those stages exist as code,
they can be reviewed in pull requests, scheduled overnight, and handed
to a colleague who did not build the original workbook. The monthly
sales reconciliation in Section 6.1 is a useful teaching case here
because it mirrors a task many students already associate with Excel,
yet it ends in a script that refuses to export if totals do not agree.

The bridge is not automatic, however. Organizations that treat
\texttt{pandas} as ``Excel but in Python'' without changing review
practices may recreate spreadsheet risk in a new medium. A notebook that
mixes exploration, hard-coded file paths, manual edits, and final
reporting in one place can be as difficult to audit as a workbook with
hidden rows and inconsistent formulas. The maturity path in Figure 7 is
therefore an organizational recommendation as much as a technical one.
Exploration can live in notebooks; repeated jobs should move toward
scripts, immutable raw data, and logged intermediate outputs as
described in Section 7.

Teams also need explicit agreements about where Excel remains
appropriate. Local review, quick what-if edits, and communication with
nontechnical stakeholders are still sensible uses of workbooks. The
problem begins when the workbook becomes the only durable record of
business logic. Table 2 is useful in these conversations because it
gives managers, instructors, and analysts a shared vocabulary for the
transition. Rather than debating whether to ``move to Python,'' a team
can ask which spreadsheet tasks are still best done by hand and which
should be expressed as repeatable table operations.

\subsection{10.2 Pandas as a Tool for Reproducible
Research}\label{pandas-as-a-tool-for-reproducible-research}

Reproducible research requires that data transformations be inspectable
and rerunnable {[}10, 11{]}. In many applied fields, the cleaning script
is not a prelude to the real analysis. It is part of the evidence that
connects raw observations to reported estimates, tables, and figures.
Reviewers and collaborators increasingly expect to see not only model
code but the path by which a messy export became an analysis-ready
table. \texttt{pandas} is well suited to that role because it keeps
intermediate results visible and because its operations map naturally to
the steps described in published methods sections.

Consider the survey workflow in Section 6.3. A methods paragraph can now
describe concrete operations: labels mapped to ordered categories, wide
responses reshaped to long format, department summaries computed at a
stated grain, and a data-quality table recorded before inference. Those
steps are not cosmetic details. They determine sample size, missingness,
and the interpretation of group means. When they live only in manual
spreadsheet edits, replication becomes guesswork. When they live in a
script with logged row counts, another researcher can compare their own
export against the same rules.

The benefit is therefore not merely automation. It is standardization of
assumptions. When a collaborator receives a cleaned \texttt{DataFrame}
without the code that produced it, they inherit hidden choices about
duplicate keys, subtotal rows, timezone parsing, and treatment of blank
strings. When they receive the script, they can challenge those choices
directly. For team-based research in public policy, health services,
education, and social science---fields where data often arrive as
administrative extracts or partner spreadsheets---this transparency can
matter as much as the estimator chosen at the end of the pipeline.

Reproducibility also depends on environment management. \texttt{pandas}
behavior can change across versions, especially around parsing, nullable
dtypes, categoricals, and copy semantics {[}1, 17{]}. A script that ran
correctly under one version may warn or fail under another if
dependencies are left floating. Research groups that publish
supplementary materials should therefore treat environment files, raw
data checksums, and processing scripts as part of the scholarly record
rather than as informal attachments. That standard is not unique to
\texttt{pandas}, but the library's centrality in Python data work makes
these habits especially important.

Notebooks remain valuable for exploration and communication {[}13{]}.
They are less dependable as the sole archive of a production pipeline
because out-of-order execution and hidden state can separate displayed
code from actual results. The examples in this paper therefore emphasize
scripts and functions for work that must be repeated on new waves of
data or reviewed by people who were not present when the first draft was
written. In practice, a healthy project often uses both modes: notebooks
to discover anomalies and scripts to freeze the rules that should
survive discovery.

\subsection{10.3 Pandas as a Business Analytics Governance
Tool}\label{pandas-as-a-business-analytics-governance-tool}

In finance, operations, compliance, and executive reporting, the
decisive question is often whether a metric can be explained and
defended under scrutiny. Spreadsheet culture already recognizes that
requirement through sign-offs, tie-outs, variance explanations, and
manual checks against control totals. Those controls reflect real
institutional needs. The weakness is that they are often applied late in
the process, after lookups, filters, pasted values, and visually
plausible formatting have already shaped the result. By the time a
reviewer approves a final tab, the path from source data to displayed
number may be difficult to reconstruct.

\texttt{pandas} allows the same institutional care to be applied earlier
and more systematically. Assertions, reconciliation checks, and
validated joins convert informal vigilance into explicit rules. The
monthly sales example in Section 6.1 is representative: detail revenue
must equal summary revenue before any Excel report is exported. That
pattern resembles month-end close in accounting, where statements are
not released until ledgers agree. Customer complaint monitoring in
Section 6.4 applies the same logic to service operations by defining
overdue cases and summarizing resolution times at a stated grain. The
SQL-to-pandas workflow in Section 6.5 shows that validation need not
disappear when authoritative data live in a database.

None of these examples require large platform investments. They require
that business rules be written where reviewers can see them and that
outputs be blocked when rules fail. That is a different culture from
``fix it in the sheet before the meeting,'' but it is not a different
culture from serious financial control. It relocates the control point
from the presentation layer to the transformation layer described in
Figure 1.

\texttt{pandas} does not guarantee correctness. A wrong rule, executed
consistently, will reproduce the same wrong answer every month with
impressive reliability. Governance still depends on domain expertise,
access control, ownership of scripts, and periodic review of
assumptions. What changes is the focal point of review. Instead of
asking only whether a final cell looks reasonable, reviewers can ask
whether join keys were validated, whether subtotals were removed,
whether identifiers were read with the correct dtype, and whether the
script version matches the period under audit. Table 3 and Table 9
collect the practices that make this shift workable for teams without
dedicated data-engineering staff.

There is also a cultural dimension that deserves emphasis. Analysts
moving from Excel to code sometimes fear they will lose contact with
business stakeholders. In the workflows presented here, the opposite
often happens. When assumptions are explicit, conversations about a
metric can center on definitions, exclusions, and edge cases instead of
on locating the latest workbook on a shared drive. Excel can remain the
format in which results are discussed. The difference is that the
workbook is generated after validation rather than serving as the only
place where logic lives. For many organizations, that distinction is the
practical meaning of analytics maturity.

\subsection{10.4 Pandas as a Teaching Tool for Computational
Thinking}\label{pandas-as-a-teaching-tool-for-computational-thinking}

For business analytics students, \texttt{pandas} teaches column-oriented
thinking, data grain, and pipeline decomposition without beginning from
infrastructure topics that can feel distant from entry-level roles. A
student who understands why \texttt{merge(validate="m:1")} matters has
learned something about relational cardinality that will transfer to
SQL, dimensional modeling, and later data engineering courses. A student
who reshapes survey data from wide to long format has met tidy data
principles {[}6{]} in a concrete task rather than as an abstract slogan.
These lessons are difficult to teach if the curriculum jumps directly
from Excel to model APIs.

Instruction can build on spreadsheet experience rather than compete with
it. Table 2 provides a concise mapping for the first weeks of a course:
VLOOKUP becomes \texttt{merge()}, pivot tables become \texttt{groupby()}
and \texttt{pivot\_table()}, and informal consistency checks become
assertions. Section 8 turns common migration mistakes into teaching
moments. Students who have lost leading zeros in a CSV export or
double-counted subtotal rows in a pivot are ready to understand why
dtype declarations and row filters belong in code. Appendix B then
supplies short patterns students can adapt in labs, projects, and
capstone work without writing every pipeline from a blank file.

There is a real pedagogical trade-off. \texttt{pandas} syntax is not
minimal, and tracebacks can intimidate beginners who are still building
Python confidence. Courses that race to machine learning estimators
without establishing table manipulation fundamentals often graduate
students who can call APIs but cannot diagnose dirty keys, unreconciled
totals, or leakage in feature construction. A stronger sequence treats
\texttt{pandas} as the place where data quality, reshaping, and
validation are learned before modeling. Section 6 offers
assignment-ready scenarios: monthly sales close, messy intake forms,
survey summaries, complaint dashboards, time-series monitoring, and
model preparation with explicit feature columns.

Teaching these workflows also creates space to discuss professional
practice. Students learn not only how to compute a metric, but how to
document it, rerun it on new data, store raw files unchanged, and
explain results to someone who will never open the script. Those habits
correspond to what employers often describe as reliability rather than
brilliance: less dependence on one analyst's heroics in a workbook, more
confidence that a pipeline can be inherited by another person next
quarter. Pairing \texttt{pandas} labs with short written
reflections---what grain does this table have, what could go wrong in
the join, what would you check before sending the report---helps connect
code exercises to workplace judgment.

\subsection{10.5 When to Move Beyond
Pandas}\label{when-to-move-beyond-pandas}

\texttt{pandas} is not the right tool for every scale or production
requirement. Its core strength is expressive in-memory transformation of
tables that fit comfortably on one machine with reasonable runtime. When
volume, concurrency, latency, or schema enforcement outgrow that
setting, forcing every task through \texttt{pandas} can produce fragile
scripts, long runtimes, or silent memory failures. Recognizing that
boundary is part of competent use rather than an admission of failure.

Very large datasets, concurrent writers, low-latency serving, and strict
schema contracts may require SQL warehouses, DuckDB, Polars, Dask,
Spark, or cloud-native pipelines {[}9, 17, 18{]}. The pattern we
advocate is progressive escalation. A team can begin with the practices
emphasized throughout this paper---immutable raw data, explicit dtypes,
validated joins, reconciliation checks, and scripted outputs---then
profile where time and memory are actually spent. If keyed aggregation
on millions of rows dominates runtime, pushing work into SQL or DuckDB
may preserve the same logic while reducing data movement. If batch jobs
outgrow a single host, groupby and validation rules developed in
\texttt{pandas} can often be translated into partition-aware jobs rather
than abandoned.

What should survive escalation is not necessarily the exact API call,
but the workflow discipline in Table 3. Ingestion rules, join
validations, reconciliation assertions, and separation of raw from
processed data remain meaningful whether execution happens on a laptop,
in a warehouse, or on a cluster. Teams that purchase larger systems
without preserving those controls frequently scale errors faster than
they scale insight. The technology changes; the need for explicit grain
and defensible totals does not.

Tool choice also depends on organizational context. A regional
nonprofit, a university research lab, and a multinational finance group
face different constraints on staffing, licensing, training, and
operations. For many applied analytics units, the hybrid architecture in
Figure 8 is sufficient for years: databases and exports as sources,
\texttt{pandas} for governed transformation, Excel or dashboards for
delivery. Larger organizations may later embed the same rules in
orchestration tools, warehouse transformations, or quality frameworks.
In both settings, \texttt{pandas} is often where analysts first
formalize data-quality logic that later becomes an institutional
standard.

Table 7 summarizes typical escalation triggers. Interpreting that table
should not feel like a mandate to abandon the library at the first sign
of slowness. It is a guide to recognizing when the same workflow
patterns should be executed on a different engine. The goal is
dependable tabular work at the scale the organization actually faces,
not maximal tooling complexity.

\subsection{10.6 Implications for
Practice}\label{implications-for-practice}

Taken together, the sections above suggest a practical research and
teaching agenda rather than a single prescription. For practitioners,
the immediate step is usually not a platform migration but a workflow
audit: which recurring reports still depend on manual workbook edits,
which joins lack validation, and which outputs are exported before
totals are reconciled. For instructors, the step is often curricular:
place table manipulation, validation, and reproducible project structure
before advanced modeling topics. For managers, the step is governance:
treat critical \texttt{pandas} scripts with the same seriousness as
critical spreadsheets, including review, ownership, and version history.

The examples and appendices in this paper are intentionally modular so
they can support those steps. A team can adopt one validated join
pattern, one reconciliation check, or one folder layout without
accepting every recommendation at once. That incremental path matches
how spreadsheet-heavy organizations actually change. \texttt{pandas}
succeeds in those settings when it reduces fragility without demanding
that analysts become a different kind of professional overnight. Its
role is to make good analytical habits easier to repeat.

\textbf{Table 7.} Tool escalation beyond in-memory pandas

{\def\LTcaptype{none} 
\begin{longtable}[]{@{}
  >{\raggedright\arraybackslash}p{(\linewidth - 6\tabcolsep) * \real{0.2500}}
  >{\raggedright\arraybackslash}p{(\linewidth - 6\tabcolsep) * \real{0.2500}}
  >{\raggedright\arraybackslash}p{(\linewidth - 6\tabcolsep) * \real{0.2500}}
  >{\raggedright\arraybackslash}p{(\linewidth - 6\tabcolsep) * \real{0.2500}}@{}}
\toprule\noalign{}
\begin{minipage}[b]{\linewidth}\raggedright
Trigger
\end{minipage} & \begin{minipage}[b]{\linewidth}\raggedright
Typical symptom
\end{minipage} & \begin{minipage}[b]{\linewidth}\raggedright
Escalation option
\end{minipage} & \begin{minipage}[b]{\linewidth}\raggedright
What transfers from pandas workflows
\end{minipage} \\
\midrule\noalign{}
\endhead
\bottomrule\noalign{}
\endlastfoot
Data exceeds RAM & Kernel OOM, swapping & SQL pushdown, DuckDB, chunking
& Same validation rules in SQL or chunked loops \\
Slow joins on millions of rows & Multi-minute merges & Warehouse SQL,
Polars & Join keys and \texttt{validate} logic as SQL constraints \\
Need concurrent writers & File locking, overwrites & Database tables &
\texttt{to\_sql} with governed schemas \\
Scheduled production jobs & Notebook failures & Airflow / scripts / dbt
& Functions extracted from \texttt{src/} \\
Strict schema contracts & Silent dtype drift & dbt, Great Expectations &
Assertions become tests \\
Distributed batch & Single-machine ceiling & Dask, Spark &
Partition-aware groupby patterns \\
\end{longtable}
}

\section{11. Limitations}\label{limitations}

Several limitations should be explicit. Table 8 complements the
escalation guidance in Table 7.

\textbf{Table 8.} Limitations and mitigations for pandas workflows

{\def\LTcaptype{none} 
\begin{longtable}[]{@{}
  >{\raggedright\arraybackslash}p{(\linewidth - 4\tabcolsep) * \real{0.3333}}
  >{\raggedright\arraybackslash}p{(\linewidth - 4\tabcolsep) * \real{0.3333}}
  >{\raggedright\arraybackslash}p{(\linewidth - 4\tabcolsep) * \real{0.3333}}@{}}
\toprule\noalign{}
\begin{minipage}[b]{\linewidth}\raggedright
Limitation area
\end{minipage} & \begin{minipage}[b]{\linewidth}\raggedright
Manifestation
\end{minipage} & \begin{minipage}[b]{\linewidth}\raggedright
Mitigation
\end{minipage} \\
\midrule\noalign{}
\endhead
\bottomrule\noalign{}
\endlastfoot
Memory & OOM on wide merges & Chunking, SQL pushdown, column pruning \\
Performance & Slow groupby at scale & Polars, DuckDB, warehouse SQL \\
Version sensitivity & Parsing / CoW behavior changes & Pin versions,
regression tests \\
Notebooks & Out-of-order execution & Scripts,
\texttt{nbconvert\ -\/-execute} \\
Weak schemas & Silent coercion & Assertions, Pandera, warehouse types \\
Governance & Leaked exports / logs & Access controls on
\texttt{reports/} \\
Excel I/O & Lost macros / complex layout & Separate raw data from
presentation sheets \\
Training cost & Misencoded business rules & Code review, pair
analytics-engineering \\
\end{longtable}
}

\textbf{Memory constraints.} \texttt{pandas} is primarily in-memory.
Wide merges, sorts, and chained copies on multi-gigabyte tables can
exhaust RAM.

\textbf{Performance on large data.} Operations that are interactive on
thousands of rows may be unacceptable on hundreds of millions without
chunking or a columnar/out-of-core engine.

\textbf{Version-sensitive behavior.} Parsing, nullable dtypes,
categoricals, time zones, and Copy-on-Write semantics evolve across
releases {[}1{]}. Long-lived projects need pinned versions and
regression checks.

\textbf{Notebook reproducibility risks.} Informal notebooks with
out-of-order execution undermine the reproducibility \texttt{pandas} can
otherwise support {[}12, 13{]}.

\textbf{Weak typing compared with schema-first systems.} Unlike dbt- or
warehouse-enforced schemas, \texttt{pandas} will happily add columns or
silently coerce types until an assertion fails.

\textbf{Data governance concerns.} Scripts can leak sensitive data
through exports, logs, or unsecured paths. Governance policies must
cover code repositories and output folders, not only databases.

\textbf{Excel interoperability limits.} \texttt{read\_excel} and
\texttt{to\_excel} do not preserve every Excel feature---macros, certain
formats, and complex layouts may be lost. Presentation-heavy workbooks
may still require manual finishing.

\textbf{Training and code review overhead.} Teams need enough Python
literacy to read pipelines and enough domain review to validate business
rules encoded in code.

\section{12. Practical Recommendations}\label{practical-recommendations}

Table 9 consolidates the practices emphasized throughout the paper. Each
row links a recommendation to a concrete pandas habit and the problem it
mitigates.

\textbf{Table 9.} Practical recommendations for governed pandas
workflows

{\def\LTcaptype{none} 
\begin{longtable}[]{@{}
  >{\raggedright\arraybackslash}p{(\linewidth - 6\tabcolsep) * \real{0.2500}}
  >{\raggedright\arraybackslash}p{(\linewidth - 6\tabcolsep) * \real{0.2500}}
  >{\raggedright\arraybackslash}p{(\linewidth - 6\tabcolsep) * \real{0.2500}}
  >{\raggedright\arraybackslash}p{(\linewidth - 6\tabcolsep) * \real{0.2500}}@{}}
\toprule\noalign{}
\begin{minipage}[b]{\linewidth}\raggedright
Recommendation
\end{minipage} & \begin{minipage}[b]{\linewidth}\raggedright
Why it matters
\end{minipage} & \begin{minipage}[b]{\linewidth}\raggedright
Example pandas practice
\end{minipage} & \begin{minipage}[b]{\linewidth}\raggedright
Problem mitigated
\end{minipage} \\
\midrule\noalign{}
\endhead
\bottomrule\noalign{}
\endlastfoot
Preserve raw data & Provenance and audit trails & Copy files to
\texttt{data/raw/}; never edit in place & Irreversible manual edits \\
Use explicit dtypes & Prevents silent coercion & Declare ID columns as
strings & Lost leading zeros, bad joins \\
Validate joins & Duplicate keys inflate totals & Validate merge
cardinality & Inflated revenue or headcount \\
Reconcile totals & Catches grain mistakes & Compare detail and summary
totals & Misleading management reports \\
Keep business rules in code & Rules become reviewable & Named functions
in \texttt{src/cleaning.py} & Hidden formula drift \\
Export Excel only after validation & Presentation should not precede QA
& Write reports last & Publishing known-bad tables \\
Use version control & Shows when logic changed & Git for \texttt{src/},
\texttt{tests/}, env files & Untracked analyst tweaks \\
Pin dependencies & Results stay reproducible & \texttt{requirements.txt}
with exact versions & ``Works on my machine'' failures \\
Use scripts for repeated workflows & Notebooks are harder to schedule &
Scripted entry point & Out-of-order notebook runs \\
Escalate beyond pandas when needed & In-memory tools have ceilings &
Move heavy joins to SQL/DuckDB/Polars & OOM errors, slow jobs \\
\end{longtable}
}

\section{13. Conclusion}\label{conclusion}

\texttt{pandas} is a foundational tool for modern data analysis because
it meets analysts where much tabular work begins: spreadsheets. Its
importance is not only technical. It supports a professional transition
from manual, cell-oriented analysis to reproducible, inspectable, and
reusable workflows.

For computational research, \texttt{pandas} provides the tabular layer
needed for cleaning, reshaping, exploratory analysis, and integration
with statistical and machine learning libraries. For business analytics,
it provides a path from Excel-centered reporting to governed pipelines
without eliminating Excel as a communication format. The strongest
\texttt{pandas} workflows combine practical data manipulation with
explicit validation. They make assumptions visible, reduce manual
repetition, and create a durable record of how results were produced.

The future of data work will remain heterogeneous. Spreadsheets,
databases, notebooks, scripts, dashboards, and machine learning systems
will continue to coexist. \texttt{pandas} is valuable because it can
connect these systems while keeping the analytical logic close to the
data and clear enough to review (Figure 8). It does not require
abandoning spreadsheets; rather, it provides a reproducible
transformation layer between raw data and decisions (Figure 1).

\section{Code and Data Availability}\label{code-and-data-availability}

The examples in this paper are illustrative and self-contained at the
code-snippet level, but they refer to representative input files such as
\texttt{monthly\_sales.xlsx}, \texttt{employee\_survey.csv},
\texttt{complaints.xlsx}, and SQL database connections to mirror common
business and research workflows. These files are not distributed as a
companion dataset. The code blocks were checked for Python syntax, but
examples that depend on external workbooks, databases, or optional
packages such as \texttt{xlsxwriter}, \texttt{sqlalchemy}, or
scikit-learn require corresponding local inputs and dependencies before
they can be executed end to end.

\section{Appendix A. Excel-to-Pandas Mini
Cookbook}\label{appendix-a.-excel-to-pandas-mini-cookbook}

\subsection{Filter Rows}\label{filter-rows}

\begin{Shaded}
\begin{Highlighting}[]
\NormalTok{west\_large\_orders }\OperatorTok{=}\NormalTok{ orders.loc[}
\NormalTok{    (orders[}\StringTok{"region"}\NormalTok{].eq(}\StringTok{"West"}\NormalTok{)) }\OperatorTok{\&}\NormalTok{ (orders[}\StringTok{"revenue"}\NormalTok{].ge(}\DecValTok{500}\NormalTok{))}
\NormalTok{]}
\end{Highlighting}
\end{Shaded}

\subsection{Create a Conditional
Column}\label{create-a-conditional-column}

\begin{Shaded}
\begin{Highlighting}[]
\NormalTok{orders }\OperatorTok{=}\NormalTok{ orders.assign(}
\NormalTok{    order\_size}\OperatorTok{=}\KeywordTok{lambda}\NormalTok{ d: pd.cut(}
\NormalTok{        d[}\StringTok{"revenue"}\NormalTok{],}
\NormalTok{        bins}\OperatorTok{=}\NormalTok{[}\DecValTok{0}\NormalTok{, }\DecValTok{100}\NormalTok{, }\DecValTok{500}\NormalTok{, }\BuiltInTok{float}\NormalTok{(}\StringTok{"inf"}\NormalTok{)],}
\NormalTok{        labels}\OperatorTok{=}\NormalTok{[}\StringTok{"Small"}\NormalTok{, }\StringTok{"Medium"}\NormalTok{, }\StringTok{"Large"}\NormalTok{],}
\NormalTok{        include\_lowest}\OperatorTok{=}\VariableTok{True}\NormalTok{,}
\NormalTok{    )}
\NormalTok{)}
\end{Highlighting}
\end{Shaded}

\subsection{Remove Duplicate Customer
Records}\label{remove-duplicate-customer-records}

\begin{Shaded}
\begin{Highlighting}[]
\NormalTok{customers\_deduped }\OperatorTok{=}\NormalTok{ (}
\NormalTok{    customers}
\NormalTok{    .sort\_values([}\StringTok{"customer\_id"}\NormalTok{, }\StringTok{"customer\_name"}\NormalTok{])}
\NormalTok{    .drop\_duplicates(subset}\OperatorTok{=}\NormalTok{[}\StringTok{"customer\_id"}\NormalTok{], keep}\OperatorTok{=}\StringTok{"first"}\NormalTok{)}
\NormalTok{)}
\end{Highlighting}
\end{Shaded}

\subsection{Find Unmatched Lookup
Keys}\label{find-unmatched-lookup-keys}

\begin{Shaded}
\begin{Highlighting}[]
\NormalTok{lookup\_check }\OperatorTok{=}\NormalTok{ orders.merge(}
\NormalTok{    customers[[}\StringTok{"customer\_id"}\NormalTok{]],}
\NormalTok{    on}\OperatorTok{=}\StringTok{"customer\_id"}\NormalTok{,}
\NormalTok{    how}\OperatorTok{=}\StringTok{"left"}\NormalTok{,}
\NormalTok{    indicator}\OperatorTok{=}\VariableTok{True}\NormalTok{,}
\NormalTok{)}

\NormalTok{unmatched }\OperatorTok{=}\NormalTok{ lookup\_check.loc[}
\NormalTok{    lookup\_check[}\StringTok{"\_merge"}\NormalTok{].eq(}\StringTok{"left\_only"}\NormalTok{),}
    \StringTok{"customer\_id"}\NormalTok{,}
\NormalTok{].drop\_duplicates()}
\end{Highlighting}
\end{Shaded}

\subsection{Convert Wide Monthly Columns to Long
Format}\label{convert-wide-monthly-columns-to-long-format}

\begin{Shaded}
\begin{Highlighting}[]
\NormalTok{wide\_budget }\OperatorTok{=}\NormalTok{ pd.DataFrame(\{}
    \StringTok{"department"}\NormalTok{: [}\StringTok{"Sales"}\NormalTok{, }\StringTok{"Finance"}\NormalTok{],}
    \StringTok{"2025{-}01"}\NormalTok{: [}\DecValTok{100000}\NormalTok{, }\DecValTok{70000}\NormalTok{],}
    \StringTok{"2025{-}02"}\NormalTok{: [}\DecValTok{110000}\NormalTok{, }\DecValTok{72000}\NormalTok{],}
\NormalTok{\})}

\NormalTok{long\_budget }\OperatorTok{=}\NormalTok{ wide\_budget.melt(}
\NormalTok{    id\_vars}\OperatorTok{=}\StringTok{"department"}\NormalTok{,}
\NormalTok{    var\_name}\OperatorTok{=}\StringTok{"month"}\NormalTok{,}
\NormalTok{    value\_name}\OperatorTok{=}\StringTok{"budget"}\NormalTok{,}
\NormalTok{).assign(month}\OperatorTok{=}\KeywordTok{lambda}\NormalTok{ d: pd.PeriodIndex(d[}\StringTok{"month"}\NormalTok{], freq}\OperatorTok{=}\StringTok{"M"}\NormalTok{))}
\end{Highlighting}
\end{Shaded}

\subsection{Reconcile Detail to
Summary}\label{reconcile-detail-to-summary}

\begin{Shaded}
\begin{Highlighting}[]
\NormalTok{detail\_total }\OperatorTok{=}\NormalTok{ enriched[}\StringTok{"revenue"}\NormalTok{].}\BuiltInTok{sum}\NormalTok{()}
\NormalTok{summary\_total }\OperatorTok{=}\NormalTok{ monthly\_metrics[}\StringTok{"revenue"}\NormalTok{].}\BuiltInTok{sum}\NormalTok{()}

\ControlFlowTok{if} \KeywordTok{not}\NormalTok{ np.isclose(detail\_total, summary\_total):}
    \ControlFlowTok{raise} \PreprocessorTok{ValueError}\NormalTok{(}
        \SpecialStringTok{f"Revenue mismatch: detail=}\SpecialCharTok{\{}\NormalTok{detail\_total}\SpecialCharTok{\}}\SpecialStringTok{, summary=}\SpecialCharTok{\{}\NormalTok{summary\_total}\SpecialCharTok{\}}\SpecialStringTok{"}
\NormalTok{    )}
\end{Highlighting}
\end{Shaded}

\section{Appendix B. Reusable Pandas Recipes for Research and Business
Analytics}\label{appendix-b.-reusable-pandas-recipes-for-research-and-business-analytics}

The following snippets are intended for direct reuse in teaching and
practice.

\subsection{B.1 Standardizing Column
Names}\label{b.1-standardizing-column-names}

\begin{Shaded}
\begin{Highlighting}[]
\NormalTok{df }\OperatorTok{=}\NormalTok{ df.rename(columns}\OperatorTok{=}\KeywordTok{lambda}\NormalTok{ c: }\BuiltInTok{str}\NormalTok{(c).strip().lower().replace(}\StringTok{" "}\NormalTok{, }\StringTok{"\_"}\NormalTok{).replace(}\StringTok{"\%"}\NormalTok{, }\StringTok{"pct"}\NormalTok{))}
\end{Highlighting}
\end{Shaded}

\subsection{B.2 Reading Excel While Preserving ID Columns as
Strings}\label{b.2-reading-excel-while-preserving-id-columns-as-strings}

\begin{Shaded}
\begin{Highlighting}[]
\NormalTok{df }\OperatorTok{=}\NormalTok{ pd.read\_excel(}\StringTok{"file.xlsx"}\NormalTok{, dtype}\OperatorTok{=}\NormalTok{\{}\StringTok{"order\_id"}\NormalTok{: }\StringTok{"string"}\NormalTok{, }\StringTok{"customer\_id"}\NormalTok{: }\StringTok{"string"}\NormalTok{\})}
\end{Highlighting}
\end{Shaded}

\subsection{B.3 Cleaning Currency
Fields}\label{b.3-cleaning-currency-fields}

\begin{Shaded}
\begin{Highlighting}[]
\NormalTok{df }\OperatorTok{=}\NormalTok{ df.assign(}
\NormalTok{    amount}\OperatorTok{=}\KeywordTok{lambda}\NormalTok{ d: (}
\NormalTok{        d[}\StringTok{"amount"}\NormalTok{]}
\NormalTok{        .astype(}\StringTok{"string"}\NormalTok{)}
\NormalTok{        .}\BuiltInTok{str}\NormalTok{.replace(}\VerbatimStringTok{r"}\PreprocessorTok{[$,]}\VerbatimStringTok{"}\NormalTok{, }\StringTok{""}\NormalTok{, regex}\OperatorTok{=}\VariableTok{True}\NormalTok{)}
\NormalTok{        .astype(}\StringTok{"Float64"}\NormalTok{)}
\NormalTok{    )}
\NormalTok{)}
\end{Highlighting}
\end{Shaded}

\subsection{B.4 Cleaning Percent
Fields}\label{b.4-cleaning-percent-fields}

\begin{Shaded}
\begin{Highlighting}[]
\NormalTok{df }\OperatorTok{=}\NormalTok{ df.assign(}
\NormalTok{    margin\_pct}\OperatorTok{=}\KeywordTok{lambda}\NormalTok{ d: (}
\NormalTok{        d[}\StringTok{"margin\_pct"}\NormalTok{]}
\NormalTok{        .astype(}\StringTok{"string"}\NormalTok{)}
\NormalTok{        .}\BuiltInTok{str}\NormalTok{.rstrip(}\StringTok{"\%"}\NormalTok{)}
\NormalTok{        .astype(}\StringTok{"Float64"}\NormalTok{)}
\NormalTok{        .div(}\DecValTok{100}\NormalTok{)}
\NormalTok{    )}
\NormalTok{)}
\end{Highlighting}
\end{Shaded}

\subsection{B.5 Parsing Dates Safely}\label{b.5-parsing-dates-safely}

\begin{Shaded}
\begin{Highlighting}[]
\NormalTok{df }\OperatorTok{=}\NormalTok{ df.assign(event\_date}\OperatorTok{=}\NormalTok{pd.to\_datetime(df[}\StringTok{"event\_date"}\NormalTok{], errors}\OperatorTok{=}\StringTok{"coerce"}\NormalTok{))}
\ControlFlowTok{assert}\NormalTok{ df[}\StringTok{"event\_date"}\NormalTok{].notna().}\BuiltInTok{all}\NormalTok{(), }\StringTok{"Invalid dates found"}
\end{Highlighting}
\end{Shaded}

\subsection{B.6 Finding Duplicate
Keys}\label{b.6-finding-duplicate-keys}

\begin{Shaded}
\begin{Highlighting}[]
\NormalTok{dupes }\OperatorTok{=}\NormalTok{ (}
\NormalTok{    customers}
\NormalTok{    .loc[customers[}\StringTok{"customer\_id"}\NormalTok{].duplicated(keep}\OperatorTok{=}\VariableTok{False}\NormalTok{)]}
\NormalTok{    .sort\_values(}\StringTok{"customer\_id"}\NormalTok{)}
\NormalTok{)}
\ControlFlowTok{assert}\NormalTok{ dupes.empty, dupes}
\end{Highlighting}
\end{Shaded}

\subsection{B.7 Validating Join
Relationships}\label{b.7-validating-join-relationships}

\begin{Shaded}
\begin{Highlighting}[]
\NormalTok{enriched }\OperatorTok{=}\NormalTok{ orders.merge(customers, on}\OperatorTok{=}\StringTok{"customer\_id"}\NormalTok{, how}\OperatorTok{=}\StringTok{"left"}\NormalTok{, validate}\OperatorTok{=}\StringTok{"m:1"}\NormalTok{)}
\end{Highlighting}
\end{Shaded}

\subsection{B.8 Finding Unmatched Lookup
Keys}\label{b.8-finding-unmatched-lookup-keys}

\begin{Shaded}
\begin{Highlighting}[]
\NormalTok{check }\OperatorTok{=}\NormalTok{ orders.merge(customers[[}\StringTok{"customer\_id"}\NormalTok{]], on}\OperatorTok{=}\StringTok{"customer\_id"}\NormalTok{, how}\OperatorTok{=}\StringTok{"left"}\NormalTok{, indicator}\OperatorTok{=}\VariableTok{True}\NormalTok{)}
\NormalTok{unmatched }\OperatorTok{=}\NormalTok{ check.loc[check[}\StringTok{"\_merge"}\NormalTok{].eq(}\StringTok{"left\_only"}\NormalTok{), }\StringTok{"customer\_id"}\NormalTok{].drop\_duplicates()}
\end{Highlighting}
\end{Shaded}

\subsection{B.9 Creating Data-Quality
Reports}\label{b.9-creating-data-quality-reports}

\begin{Shaded}
\begin{Highlighting}[]
\NormalTok{quality }\OperatorTok{=}\NormalTok{ pd.DataFrame(\{}
    \StringTok{"column"}\NormalTok{: df.columns,}
    \StringTok{"missing"}\NormalTok{: df.isna().}\BuiltInTok{sum}\NormalTok{().values,}
    \StringTok{"dtype"}\NormalTok{: df.dtypes.astype(}\StringTok{"string"}\NormalTok{).values,}
\NormalTok{\})}
\end{Highlighting}
\end{Shaded}

\subsection{B.10 Reconciling Detail and Summary
Tables}\label{b.10-reconciling-detail-and-summary-tables}

\begin{Shaded}
\begin{Highlighting}[]
\ImportTok{import}\NormalTok{ numpy }\ImportTok{as}\NormalTok{ np}
\ControlFlowTok{assert}\NormalTok{ np.isclose(detail[}\StringTok{"revenue"}\NormalTok{].}\BuiltInTok{sum}\NormalTok{(), summary[}\StringTok{"revenue"}\NormalTok{].}\BuiltInTok{sum}\NormalTok{())}
\end{Highlighting}
\end{Shaded}

\subsection{B.11 Converting Wide Data to Long
Data}\label{b.11-converting-wide-data-to-long-data}

\begin{Shaded}
\begin{Highlighting}[]
\NormalTok{long\_df }\OperatorTok{=}\NormalTok{ wide\_df.melt(id\_vars}\OperatorTok{=}\NormalTok{[}\StringTok{"id"}\NormalTok{], var\_name}\OperatorTok{=}\StringTok{"period"}\NormalTok{, value\_name}\OperatorTok{=}\StringTok{"value"}\NormalTok{)}
\end{Highlighting}
\end{Shaded}

\subsection{B.12 Creating Pivot-Style
Summaries}\label{b.12-creating-pivot-style-summaries}

\begin{Shaded}
\begin{Highlighting}[]
\NormalTok{pivot }\OperatorTok{=}\NormalTok{ df.pivot\_table(}
\NormalTok{    index}\OperatorTok{=}\StringTok{"month"}\NormalTok{,}
\NormalTok{    columns}\OperatorTok{=}\StringTok{"region"}\NormalTok{,}
\NormalTok{    values}\OperatorTok{=}\StringTok{"revenue"}\NormalTok{,}
\NormalTok{    aggfunc}\OperatorTok{=}\StringTok{"sum"}\NormalTok{,}
\NormalTok{    fill\_value}\OperatorTok{=}\DecValTok{0}\NormalTok{,}
\NormalTok{)}
\end{Highlighting}
\end{Shaded}

\subsection{B.13 Exporting Multi-Sheet Excel
Reports}\label{b.13-exporting-multi-sheet-excel-reports}

\begin{Shaded}
\begin{Highlighting}[]
\ControlFlowTok{with}\NormalTok{ pd.ExcelWriter(}\StringTok{"report.xlsx"}\NormalTok{, engine}\OperatorTok{=}\StringTok{"xlsxwriter"}\NormalTok{) }\ImportTok{as}\NormalTok{ writer:}
\NormalTok{    detail.to\_excel(writer, sheet\_name}\OperatorTok{=}\StringTok{"Detail"}\NormalTok{, index}\OperatorTok{=}\VariableTok{False}\NormalTok{)}
\NormalTok{    summary.to\_excel(writer, sheet\_name}\OperatorTok{=}\StringTok{"Summary"}\NormalTok{, index}\OperatorTok{=}\VariableTok{False}\NormalTok{)}
\end{Highlighting}
\end{Shaded}

\subsection{B.14 Adding Basic Workbook
Formatting}\label{b.14-adding-basic-workbook-formatting}

\begin{Shaded}
\begin{Highlighting}[]
\ControlFlowTok{with}\NormalTok{ pd.ExcelWriter(}\StringTok{"report.xlsx"}\NormalTok{, engine}\OperatorTok{=}\StringTok{"xlsxwriter"}\NormalTok{) }\ImportTok{as}\NormalTok{ writer:}
\NormalTok{    summary.to\_excel(writer, sheet\_name}\OperatorTok{=}\StringTok{"Summary"}\NormalTok{, index}\OperatorTok{=}\VariableTok{False}\NormalTok{)}
\NormalTok{    money }\OperatorTok{=}\NormalTok{ writer.book.add\_format(\{}\StringTok{"num\_format"}\NormalTok{: }\StringTok{"$\#,\#\#0"}\NormalTok{\})}
\NormalTok{    writer.sheets[}\StringTok{"Summary"}\NormalTok{].set\_column(}\StringTok{"B:B"}\NormalTok{, }\DecValTok{14}\NormalTok{, money)}
\end{Highlighting}
\end{Shaded}

\subsection{B.15 Creating Reproducible Folder
Structures}\label{b.15-creating-reproducible-folder-structures}

\begin{Shaded}
\begin{Highlighting}[]
\ImportTok{from}\NormalTok{ pathlib }\ImportTok{import}\NormalTok{ Path}
\ControlFlowTok{for}\NormalTok{ folder }\KeywordTok{in}\NormalTok{ [}\StringTok{"data/raw"}\NormalTok{, }\StringTok{"data/interim"}\NormalTok{, }\StringTok{"data/processed"}\NormalTok{, }\StringTok{"reports"}\NormalTok{, }\StringTok{"src"}\NormalTok{, }\StringTok{"tests"}\NormalTok{]:}
\NormalTok{    Path(folder).mkdir(parents}\OperatorTok{=}\VariableTok{True}\NormalTok{, exist\_ok}\OperatorTok{=}\VariableTok{True}\NormalTok{)}
\end{Highlighting}
\end{Shaded}

\subsection{B.16 Saving Intermediate
Outputs}\label{b.16-saving-intermediate-outputs}

\begin{Shaded}
\begin{Highlighting}[]
\NormalTok{df.to\_parquet(}\StringTok{"data/interim/orders\_clean.parquet"}\NormalTok{, index}\OperatorTok{=}\VariableTok{False}\NormalTok{)}
\end{Highlighting}
\end{Shaded}

\subsection{B.17 Logging Row Counts at Each Pipeline
Step}\label{b.17-logging-row-counts-at-each-pipeline-step}

\begin{Shaded}
\begin{Highlighting}[]
\ImportTok{import}\NormalTok{ logging}
\NormalTok{logging.info(}\StringTok{"after\_join: }\SpecialCharTok{\%s}\StringTok{ rows"}\NormalTok{, }\SpecialStringTok{f"}\SpecialCharTok{\{}\BuiltInTok{len}\NormalTok{(df)}\SpecialCharTok{:,\}}\SpecialStringTok{"}\NormalTok{)}
\end{Highlighting}
\end{Shaded}

\subsection{B.18 Adding Simple
Assertions}\label{b.18-adding-simple-assertions}

\begin{Shaded}
\begin{Highlighting}[]
\ControlFlowTok{assert}\NormalTok{ df[}\StringTok{"order\_id"}\NormalTok{].is\_unique}
\ControlFlowTok{assert}\NormalTok{ df[}\StringTok{"amount"}\NormalTok{].ge(}\DecValTok{0}\NormalTok{).}\BuiltInTok{all}\NormalTok{()}
\end{Highlighting}
\end{Shaded}

\subsection{B.19 Creating a Reusable Cleaning
Function}\label{b.19-creating-a-reusable-cleaning-function}

\begin{Shaded}
\begin{Highlighting}[]
\KeywordTok{def}\NormalTok{ clean\_orders(df: pd.DataFrame) }\OperatorTok{{-}\textgreater{}}\NormalTok{ pd.DataFrame:}
    \ControlFlowTok{return}\NormalTok{ (}
\NormalTok{        df.rename(columns}\OperatorTok{=}\KeywordTok{lambda}\NormalTok{ c: }\BuiltInTok{str}\NormalTok{(c).strip().lower())}
\NormalTok{        .dropna(subset}\OperatorTok{=}\NormalTok{[}\StringTok{"order\_id"}\NormalTok{])}
\NormalTok{        .assign(order\_date}\OperatorTok{=}\KeywordTok{lambda}\NormalTok{ d: pd.to\_datetime(d[}\StringTok{"order\_date"}\NormalTok{], errors}\OperatorTok{=}\StringTok{"coerce"}\NormalTok{))}
\NormalTok{    )}
\end{Highlighting}
\end{Shaded}

\subsection{B.20 Building a Simple End-to-End
Pipeline}\label{b.20-building-a-simple-end-to-end-pipeline}

\begin{Shaded}
\begin{Highlighting}[]
\KeywordTok{def}\NormalTok{ run\_pipeline(raw\_path: }\BuiltInTok{str}\NormalTok{, report\_path: }\BuiltInTok{str}\NormalTok{) }\OperatorTok{{-}\textgreater{}} \VariableTok{None}\NormalTok{:}
\NormalTok{    orders }\OperatorTok{=}\NormalTok{ pd.read\_excel(raw\_path, dtype}\OperatorTok{=}\NormalTok{\{}\StringTok{"order\_id"}\NormalTok{: }\StringTok{"string"}\NormalTok{\})}
\NormalTok{    clean }\OperatorTok{=}\NormalTok{ clean\_orders(orders)}
\NormalTok{    summary }\OperatorTok{=}\NormalTok{ clean.groupby(}
\NormalTok{        clean[}\StringTok{"order\_date"}\NormalTok{].dt.to\_period(}\StringTok{"M"}\NormalTok{)}
\NormalTok{    ).agg(revenue}\OperatorTok{=}\NormalTok{(}\StringTok{"revenue"}\NormalTok{, }\StringTok{"sum"}\NormalTok{))}
    \ControlFlowTok{assert}\NormalTok{ clean[}\StringTok{"order\_id"}\NormalTok{].is\_unique}
\NormalTok{    summary.to\_excel(report\_path)}
\end{Highlighting}
\end{Shaded}

\section{References}\label{references}

{[}1{]} The pandas development team.
\href{https://pandas.pydata.org/docs/}{\emph{pandas documentation}}.

{[}2{]} McKinney, W. (2010). Data Structures for Statistical Computing
in Python. \emph{Proceedings of the 9th Python in Science Conference},
56-61.
\href{https://doi.org/10.25080/Majora-92bf1922-00a}{doi:10.25080/Majora-92bf1922-00a}.

{[}3{]} Harris, C. R., Millman, K. J., van der Walt, S. J., et
al.~(2020). Array programming with NumPy. \emph{Nature}, 585, 357-362.
\href{https://doi.org/10.1038/s41586-020-2649-2}{doi:10.1038/s41586-020-2649-2}.

{[}4{]} Pedregosa, F., Varoquaux, G., Gramfort, A., et al.~(2011).
Scikit-learn: Machine Learning in Python. \emph{Journal of Machine
Learning Research}, 12, 2825-2830.

{[}5{]} Hunter, J. D. (2007). Matplotlib: A 2D Graphics Environment.
\emph{Computing in Science \& Engineering}, 9(3), 90-95.
\href{https://doi.org/10.1109/MCSE.2007.55}{doi:10.1109/MCSE.2007.55}.

{[}6{]} Wickham, H. (2014). Tidy Data. \emph{Journal of Statistical
Software}, 59(10), 1-23.
\href{https://doi.org/10.18637/jss.v059.i10}{doi:10.18637/jss.v059.i10}.

{[}7{]} Panko, R. R. (2008). Spreadsheet Errors: What We Know. What We
Think We Can Do.
\href{https://arxiv.org/abs/0802.3457}{arXiv:0802.3457}.

{[}8{]} Panko, R. R. (2016). What We Don't Know About Spreadsheet Errors
Today: The Facts, Why We Don't Believe Them, and What We Need to Do.
\href{https://arxiv.org/abs/1602.02601}{arXiv:1602.02601}.

{[}9{]} The pandas development team.
\href{https://pandas.pydata.org/docs/user_guide/scale.html}{\emph{Scaling
to large datasets}}.

{[}10{]} Peng, R. D. (2011). Reproducible Research in Computational
Science. \emph{Science}, 334(6060), 1226-1227.
\href{https://doi.org/10.1126/science.1213847}{doi:10.1126/science.1213847}.

{[}11{]} Sandve, G. K., Nekrutenko, A., Taylor, J., \& Hovig, E. (2013).
Ten Simple Rules for Reproducible Computational Research. \emph{PLoS
Computational Biology}, 9(10), e1003285.
\href{https://doi.org/10.1371/journal.pcbi.1003285}{doi:10.1371/journal.pcbi.1003285}.

{[}12{]} Grus, J. (2018). \emph{I Don't Like Notebooks}. JupyterCon
presentation.

{[}13{]} Rule, A., Birmingham, A., Zuniga, C., Altintas, I., \& Huang,
S. C. (2019). Ten simple rules for writing and sharing computational
analyses in Jupyter Notebooks. \emph{PLoS Computational Biology}, 15(7),
e1007007.
\href{https://doi.org/10.1371/journal.pcbi.1007007}{doi:10.1371/journal.pcbi.1007007}.

{[}14{]} The pandas development team.
\href{https://pandas.pydata.org/docs/user_guide/io.html}{\emph{IO tools
(text, CSV, HDF5, \ldots)}}.

{[}15{]} Buitinck, L., Louppe, G., Blondel, M., et al.~(2013). API
design for machine learning software: experiences from the scikit-learn
project. \emph{ECML PKDD Workshop: Languages for Data Mining and Machine
Learning}, 108-122.

{[}16{]} Powell, S. G., Baker, K. R., \& Lawson, B. (2009). A Critical
Review of the Literature on Spreadsheet Errors. \emph{Decision Support
Systems}, 47(2), 128-138.
\href{https://doi.org/10.1016/j.dss.2009.02.002}{doi:10.1016/j.dss.2009.02.002}.

{[}17{]} Richardson, D., \& pandas Contributors.
\href{https://pandas.pydata.org/docs/whatsnew/index.html}{\emph{pandas
2.0 and beyond}}.

{[}18{]} Raasveldt, M., \& Mühleisen, H. (2019). DuckDB: an Embeddable
Analytical Database. \emph{Proceedings of the 2019 International
Conference on Management of Data (SIGMOD)}, 1981-1984.
\href{https://doi.org/10.1145/3299869.3320212}{doi:10.1145/3299869.3320212}.

{[}19{]} Pimentel, J. F., Murta, L., Braganholo, V., \& Freire, J.
(2021). Understanding and improving the quality and reproducibility of
Jupyter notebooks. \emph{Empirical Software Engineering}, 26, Article
65.
\href{https://doi.org/10.1007/s10664-021-09961-9}{doi:10.1007/s10664-021-09961-9}.

{[}20{]} Wang, J., Li, L., \& Zeller, A. (2020). Assessing and restoring
reproducibility of Jupyter notebooks. \emph{Proceedings of the 35th
IEEE/ACM International Conference on Automated Software Engineering},
138-149.
\href{https://doi.org/10.1145/3324884.3416585}{doi:10.1145/3324884.3416585}.

{[}21{]} Chattopadhyay, S., Prasad, I., Henley, A. Z., Sarma, A., \&
Barik, T. (2020). What's Wrong with Computational Notebooks? Pain
Points, Needs, and Design Opportunities. \emph{Proceedings of the 2020
CHI Conference on Human Factors in Computing Systems}, 1-12.
\href{https://doi.org/10.1145/3313831.3376729}{doi:10.1145/3313831.3376729}.

{[}22{]} Samuel, S., \& Mietchen, D. (2024). Computational
reproducibility of Jupyter notebooks from biomedical publications.
\emph{GigaScience}, 13, giad113.
\href{https://doi.org/10.1093/gigascience/giad113}{doi:10.1093/gigascience/giad113}.

{[}23{]} Huang, R., Ravi, S., He, M., Tian, B., Lerner, S., \& Coblenz,
M. (2025). How Scientists Use Jupyter Notebooks: Goals, Quality
Attributes, and Opportunities. \emph{47th IEEE/ACM International
Conference on Software Engineering (ICSE)}, 1243-1255.
\href{https://doi.org/10.1109/ICSE55347.2025.00232}{doi:10.1109/ICSE55347.2025.00232}.

\end{document}